\documentclass[oneside,english,table,dvipsnames,svgnames]{aa}
\usepackage[T1]{fontenc}
\usepackage{geometry}

\usepackage{graphicx}%
\usepackage[dvipsnames]{xcolor}
\usepackage{soul}
\usepackage[unicode=true,pdfusetitle,
 bookmarks=true,bookmarksnumbered=false,bookmarksopen=false,
 breaklinks=false,pdfborder={0 0 0},pdfborderstyle={},backref=false,colorlinks=true]
 {hyperref}
\hypersetup{
 urlcolor=blue,citecolor=blue}
\usepackage{cleveref}
\usepackage{subcaption}
\usepackage{tikz}

\makeatletter

\newcommand{\ero}[0]{{\textit{eROSITA}}}

\makeatother

\begin{document}
\title{A large population of over-massive black hole quasars at z=0.3-0.8 revealed by eROSITA}

\abstract{
In most galaxies, the central black hole accounts for 
no more than a percent of the total mass in 
stars. Recently, however, extremely over-massive black holes with ratios of 
10\% have been reported in dwarf galaxies at $z<1$
and at cosmic dawn ($z>5.5$) by JWST. Both findings have been 
interpreted as signatures of the still mysterious origins of 
super-massive black holes, such that most of the black hole mass was built at birth rather than through black hole accretion. Here we show that among evolved galaxies over-massive black holes are also present, indicating that overmassive BHs are not a signature unique to black hole formation channels.
The first large-area sky survey of the eROSITA X-ray telescope on board SpectrRG identified 200 quasars by their luminous hard X-ray radiation. These signpost rapidly growing black holes. 
Complementary optical spectroscopy from the Sloan Digital Sky Survey and
archival UV to IR photometric data combined with galaxy-quasar decomposition techniques
allow us unbiased estimates of cosmological distances, black hole masses and 
host galaxy stellar masses. 
We securely identify a sample of over-massive 
black holes with BH-to-host ratios of more than 5\%, which may have undergone exponential accretion spurts lasting about a billion years.
Our survey identified a high space density of at least $4/Gpc^3$ of overmassive black holes near cosmic noon. This indicates an accretion channel disconnected from the stellar population that cause strong deviations from galaxy scaling relations. This channel is currently not part of galaxy evolution models.
The identified channel, if applicable also for the first billion years of cosmic time, can explain JWST AGN without requiring them to signify imprints of black hole seeding mechanism.
}

\author{Johannes Buchner\inst{1}\thanks{\protect\href{mailto:johannes.buchner.acad@gmx.com}{johannes.buchner.acad@gmx.com}} \and
Isabelle Gauger\inst{1} \and
Qiaoya Wu\inst{2} \and
Hattie Starck\inst{1} \and
Catarina Aydar\inst{1} \and
Yue Shen\inst{2} \and
Vardha N. Bennert\inst{3} \and
Kirpal Nandra\inst{1} \and
Sophia G. H. Waddell\inst{1,4} \and
Andrea Merloni\inst{1} \and
Mara Salvato\inst{1} \and
Roberto J. Assef\inst{5} \and
Zsofi Igo\inst{1,6} \and
Franz E. Bauer\inst{7} \and
Dong-Woo Kim\inst{8} \and
Anton M. Koekemoer\inst{9} \and
Donald P. Schneider\inst{10,11}
}

\institute{
Max Planck Institute for Extraterrestrial Physics, Giessenbachstrasse, 85748 Garching, Germany
\and
Department of Astronomy, University of Illinois at Urbana-Champaign, Urbana, IL 61801, USA
\and
Physics Department, California Polytechnic State University, 1 Grand Avenue, San Luis Obispo, CA 93407, USA
\and
Trottier Space Institute and Department of Physics, McGill University, 3600 rue University, Montreal, Quebec H3A 2T8, Canada
\and
Instituto de Estudios Astrof\'isicos, Facultad de Ingenier\'ia y Ciencias, Universidad Diego Portales, Av. Ej\'ercito Libertador 441, Santiago, Chile
\and
Exzellenzcluster ORIGINS, Boltzmannstr. 2, Garching, 85748, Germany
\and
Instituto de Alta Investigaci{\'{o}}n, Universidad de Tarapac{\'{a}}, Casilla 7D, Arica, Chile
\and
Center for Astrophysics | Harvard \& Smithsonian, 60 Garden Street, Cambridge, MA 02138, USA
\and
Space Telescope Science Institute, 3700 San Martin Drive, Baltimore, MD 21218, USA
\and
Department of Astronomy and Astrophysics, The Pennsylvania State University, University Park, PA 16802, USA
\and
Institute for Gravitation and the Cosmos, The Pennsylvania State University, University Park, PA 16802, USA
}
\date{-Received date / Accepted date}
\titlerunning{Abundance of overmassive black hole quasars}
\authorrunning{Buchner et al.}

\keywords{Galaxy evolution, black holes}

\maketitle

\section{Introduction}

The supermassive black hole (SMBH) population living at the centers of galaxies grows in mass via accretion disks that shine as Active Galactic Nuclei
\citep[AGN;][]{Burbidge1959,Burbidge1964,Lynden-Bell1969,Lynds1963,Salpeter1964,Woltjer1959,Zeldovich1964}.
However, accretion fueling is stochastic \citep{Hickox2014,Schawinski2015}, so black hole growth can lead or lag star formation.
Over-massive SMBHs, where the ratio of black hole to total stellar mass $M_{\bullet}/M_{\star}$ is 4-10 times higher than typical in quiescent galaxies, have been found in AGN samples \citep[e.g.,][]{Reines2015,Trakhtenbrot2015}.
With similar ratios, in dwarf galaxies ($M_{\star}\sim10^{9}M_{\odot}$, $z<1$) seven overmassive AGN with $M_{\bullet}\sim10^{8}M_{\odot}$ have recently been 
reported~\citep{Mezcua2023} and interpreted as dormant remnants of SMBH formation in the first half billion years of the Universe. These remnants likely evolve in isolation: The rare galaxy mergers would tend to realign systems with scaling relations \citep{Jahnke2011}.
It is now possible to start probing cosmic dawn thanks to JWST, which has revealed several exceptionally over-massive AGN at $M_{\bullet}\sim10^{7}M_{\odot}$ and $M_{\star}\sim10^{8}M_{\odot}$ \citep{Harikane2023,Pacucci2023,Maiolino2024}.
All of the above results benefit from luminous AGN emission preferentially pinpointing
more massive black holes \citep{Lauer2007}. This is in part because the luminosity distribution
at any black hole mass is steeply declining \citep{Aird2012,Bongiorno2012}, making it rare for moderately massive black holes
to reach high luminosities \citep{Lauer2007,Merloni2010}. This selection effect of luminous AGN makes them a useful tool to study extreme galaxies.
A complication in identifying over-massive SMBHs is that large samples of black hole masses are almost exclusively estimated from single-epoch spectra, a method empirically calibrated at $z<2.5$ from $\mathrm{H}\beta$ \citep{Peterson1993,Peterson2000} or $\mathrm{MgII}$
broad emission lines \citep{McLure2002}. Due to unknowns in geometry, orientation, and the virialized fraction, mass estimates carry a systematic uncertainty of about a factor of three \citep{Shen2024} that needs to be carefully considered.
Those systems effectively identified as highly over-massive black holes 
have been interpreted as being in an efficient growth phase linked to
how seed black holes become supermassive. In this work, we define over-massive black holes (OMBHs) as $M_{\bullet}/M_{\star}>5\%$, ten times above local galaxy scaling relations.

This paper is structured as follows:
In \cref{sec:data}, we present a luminous X-ray AGN selection with a remarkably high success rate in identifying OMBHs, based on high spectral completeness giving black hole mass measurements (\cref{sec:method:bhmass}) and reliable stellar mass estimation  (\cref{sec:method:stellarmass}). The OMBH sample of eight objects is characterized in \cref{sec:results}. We quantify the biases of the OMBH selection with simulations in \cref{sec:resultslauer}, including the systematic uncertainties from single-epoch black hole mass measurements. 
In \cref{sec:results:rm} we emphasize that unlike other works, our H$\beta$ black hole masses at $z=0.3-1$ lie within the distribution where single-epoch masses were calibrated in terms of moderate Eddington ratios, black hole mass and redshift.
In \cref{sec:results:lit} we compare to local galaxy samples, AGN samples at cosmic dawn, cosmic noon and the local Universe. In \cref{sec:resultsdensity}, our survey reveals the highest minimum space density of OMBHs yet inferred from AGN surveys, providing an important bridge to the local OMBH galaxy population. In \cref{sec:discussion}, we discuss our results, the reasons for the comparatively high selection efficiency (\cref{sec:discussion:selefficiency}), reliability and completeness (\cref{sec:discussion:selreliability}), interpret the comparison to cosmic dawn OMBH samples (\cref{sec:discussion:litcomparison}). Finally, \cref{sec:discussion:evolution} presents potential evolutionary tracks for becoming an OMBH at
$z=0.3-1$ and their subsequent evolution.
We assume a \cite{Planck18} cosmology and magnitudes are in AB unless stated otherwise.

\section{Sample}\label{sec:data}

We select luminous AGN from a wide-area X-ray survey conducted
by \ero{},
an imaging X-ray telescope on-board
Spectr-Roentgen-Gamma (SRG), sensitive in the 0.2-5\,keV energy range \citep{Predehl2020}.
The \ero{} full equatorial depth field (eFEDS) is a 140\,deg\texttwosuperior{}
extragalactic field surveyed by \ero{} in 2019 \citep{Brunner2021}
with rich multiwavelength photometric datasets \citep{Salvato2021}. 
The parent sample of 246 point-like X-ray sources detected in the 2.3–5 keV band contains, after removing stars, 200 AGN \citep{Nandra2025}. This parent sample
has also been the subject of optical spectroscopic follow-up campaigns with the Sloan Digital Sky Survey  \citep[SDSS;][]{Gunn2006SDSS,Smee2013SDSS,Kollmeier2017,Kollmeier2025} IV and V, yielding secure AGN identifications with high (80\%) completeness.
Most of the parent sample lies in the $z=0-1$
redshift range (88\%) and has rest-frame, absorption-corrected X-ray luminosities
of $L_{\mathrm{0.5-2keV}}=10^{43-45}\mathrm{erg/s}$.
The hard sample sources form the bulk of our selection robustness analysis.

To increase the completeness in the field, we also consider sources detected in the 
0.5-2.3\,keV eFEDS survey described in \cite{Liu2021a} and \cite{Salvato2021}. 
We analyse these eFEDS-main sources in the same fashion as the eFEDS-hard sources. 
However, for modeling the selection function, the hard sample is preferable due to its reduced sensitivity to absorption and to additional soft X-ray emission components \citep{Waddell2024SX}.
Throughout, we plot eFEDS-hard sources in purple and eFEDS-main sources in pink, and use circles for measurements and triangles for upper or lower limits.

\section{Methods}\label{sec:method}

\subsection{Black hole mass estimate}\label{sec:method:bhmass}

\begin{figure*}
\begin{centering}
\includegraphics[width=0.85\textwidth]{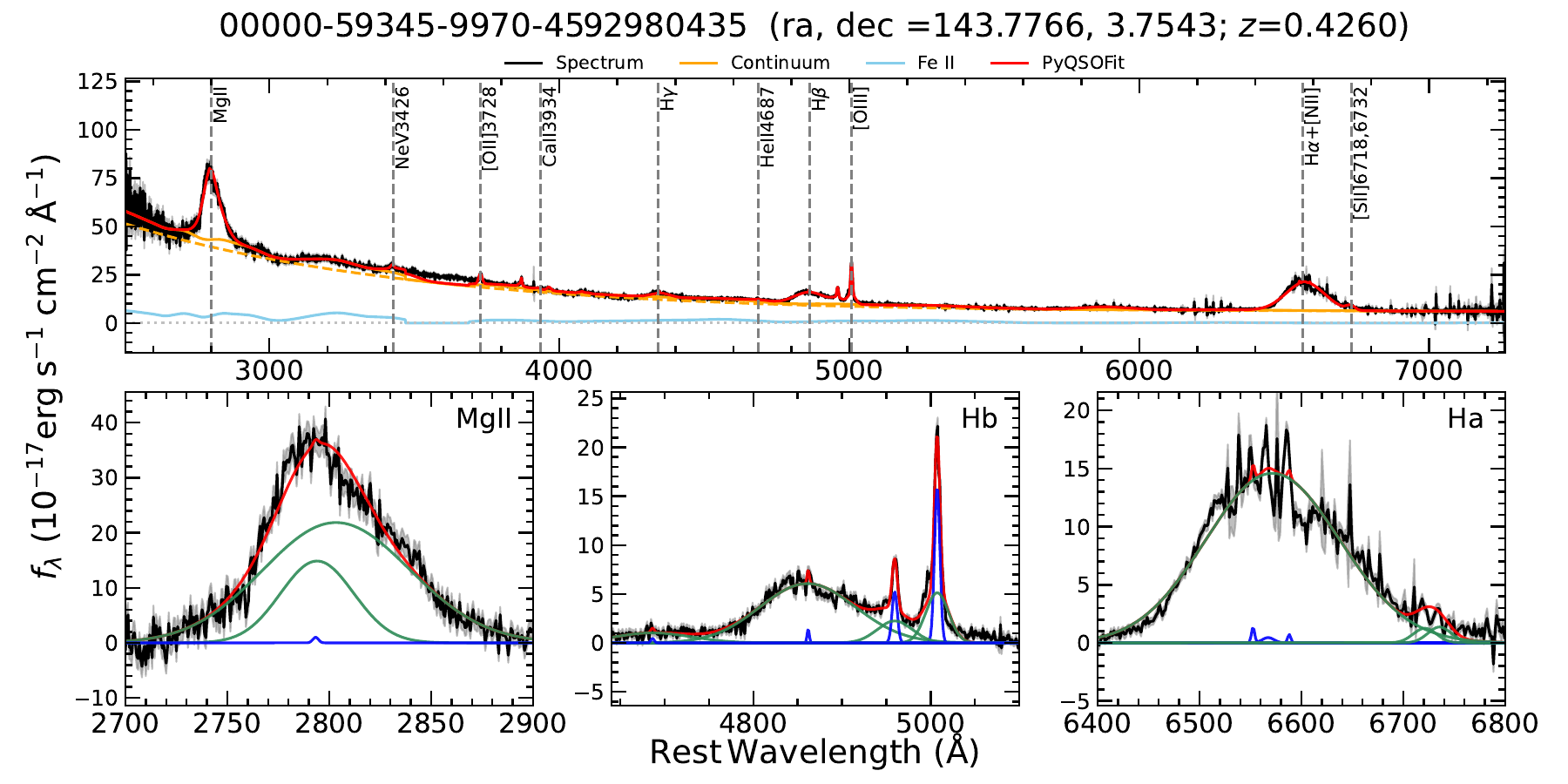}
\par\end{centering}
\begin{centering}
\includegraphics[width=0.85\textwidth]{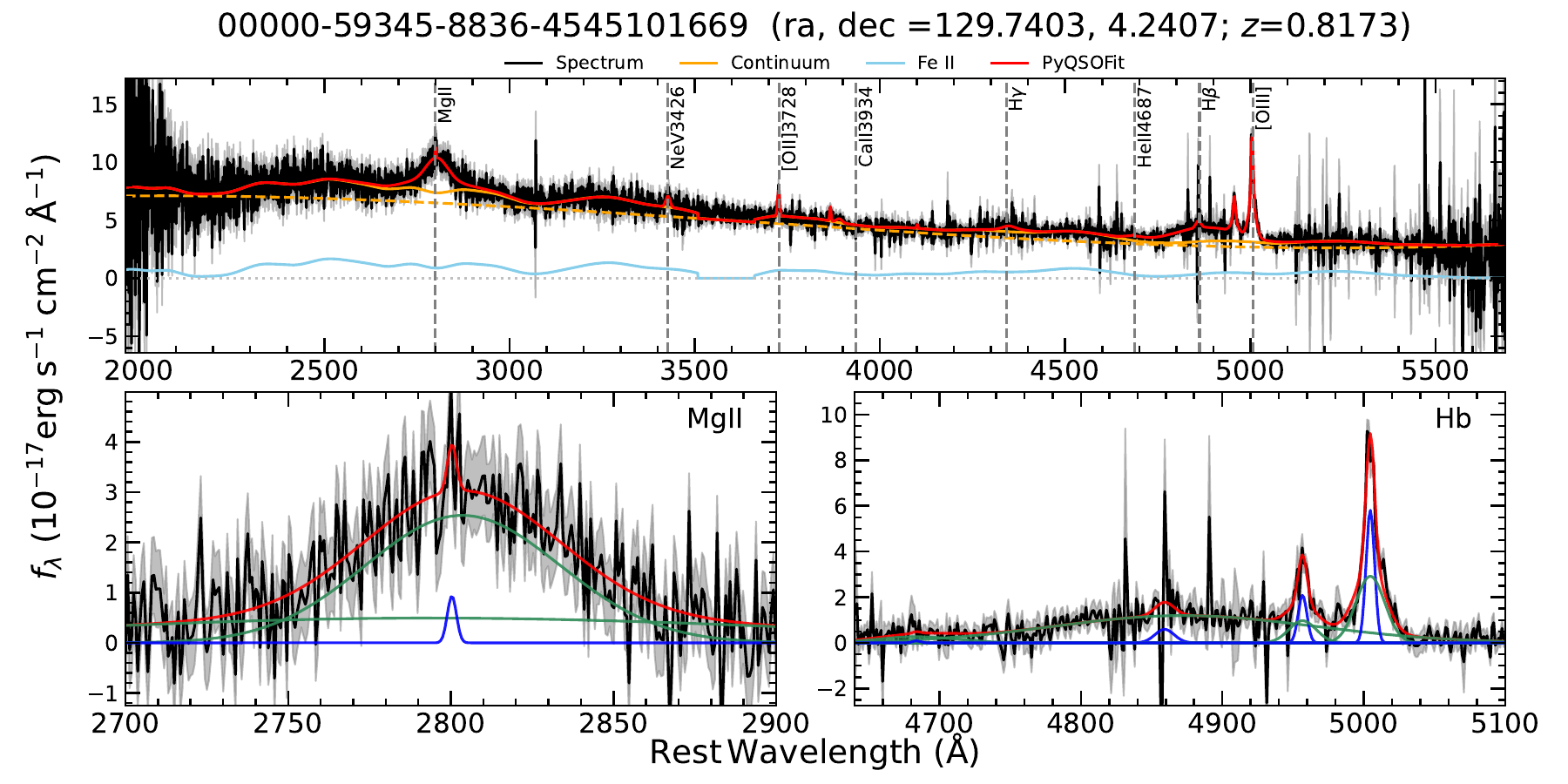}
\par\end{centering}
\begin{centering}
\includegraphics[width=0.85\textwidth]{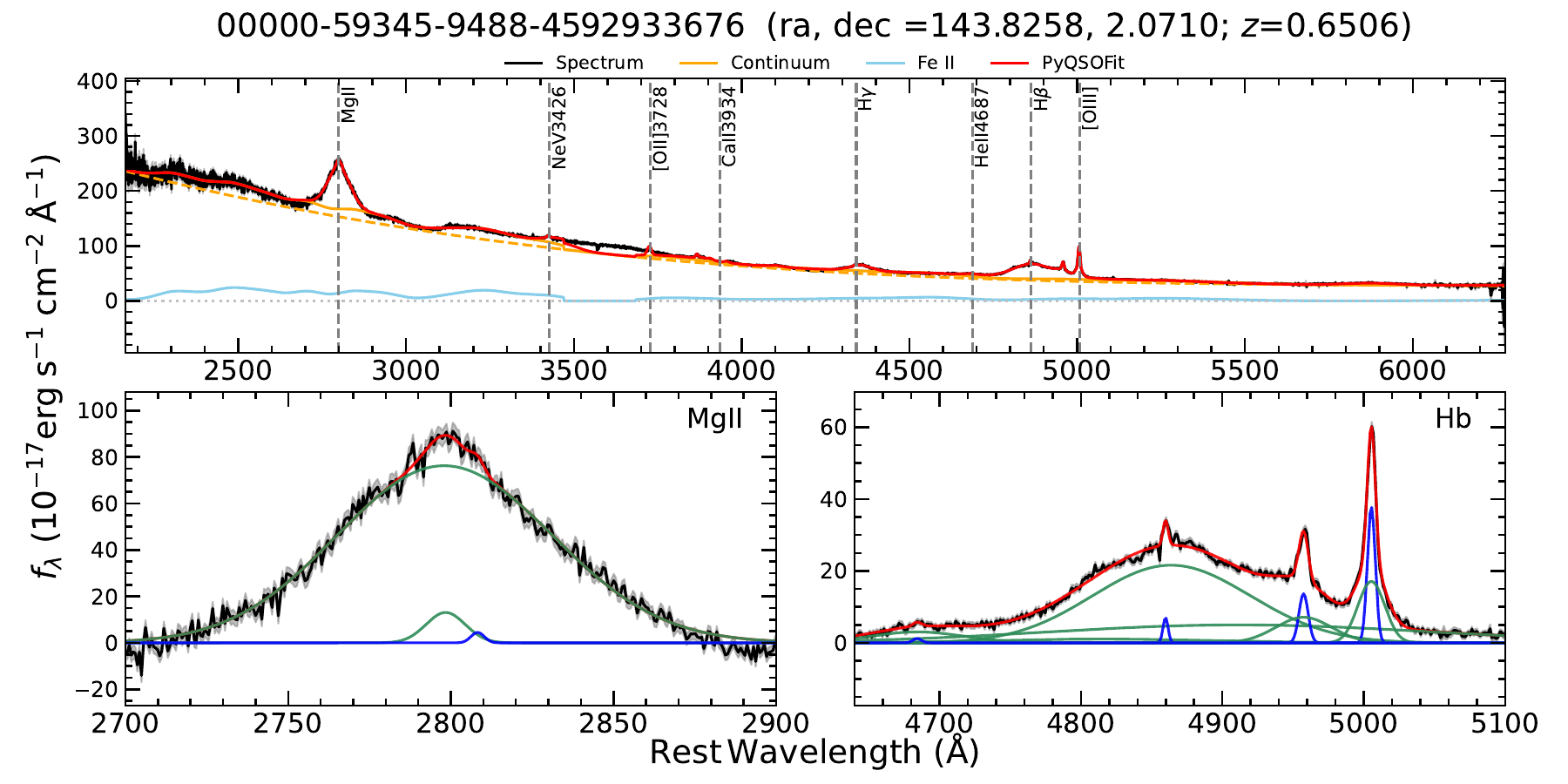}
\par\end{centering}
\caption{Optical spectral fits of ID 133 (top panel), 1136 (middle panel) and
32 (bottom panel), our three OMBH quasars from the eFEDS-hard sample. 
The title gives the SDSS ID, position and redshift assumed during fitting.
After continuum modeling (upper inset), each line complex (inset) is modeled with up to three broad components, which are combined when measuring the total full width half maximum.
\protect\label{fig:optspec}}
\end{figure*}

\begin{figure*}
\begin{centering}
\includegraphics[width=\textwidth,trim={0 0 3cm 1.5cm},clip]{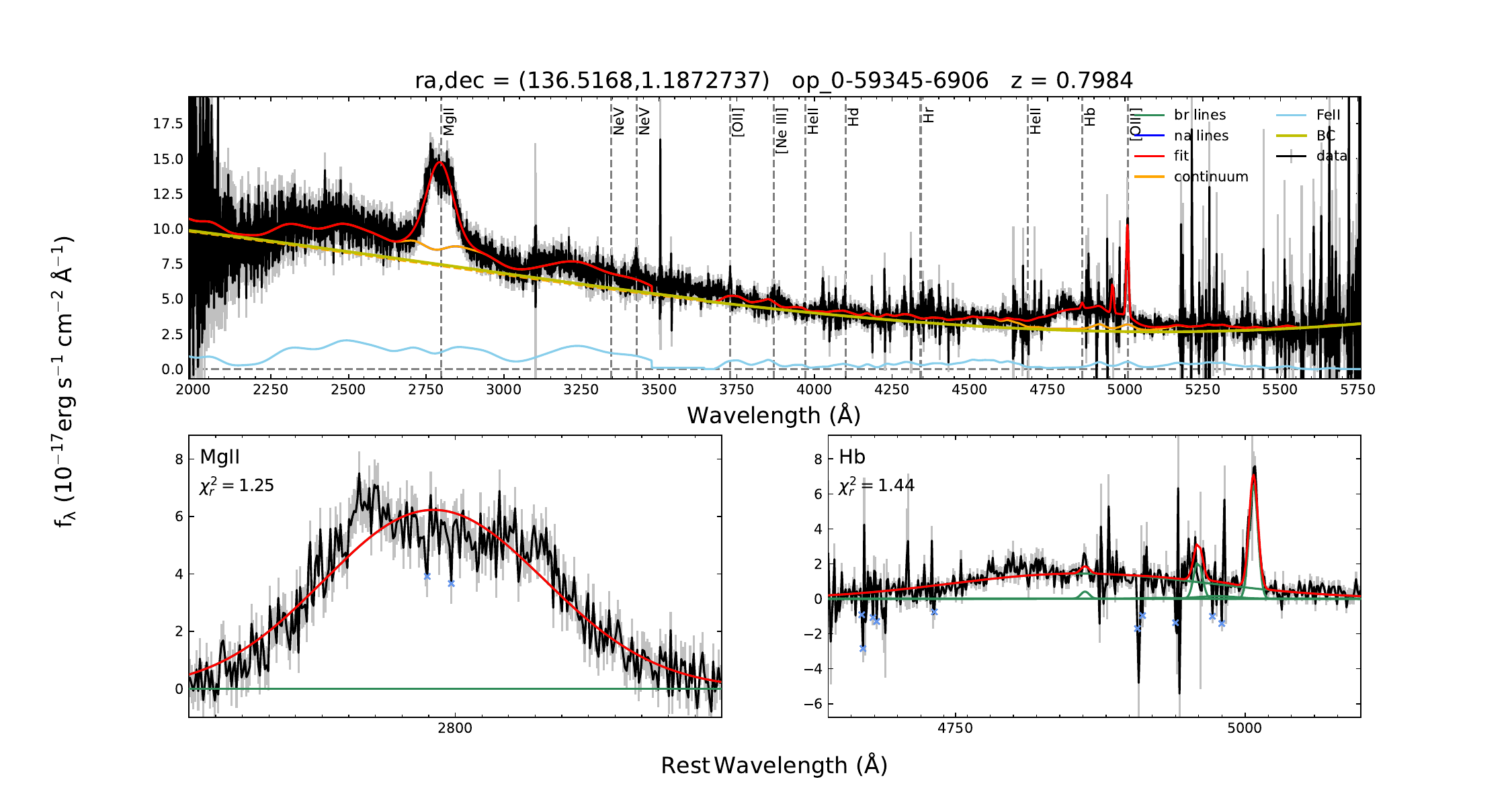}
\includegraphics[width=\textwidth,trim={0 0 3cm 1.5cm},clip]{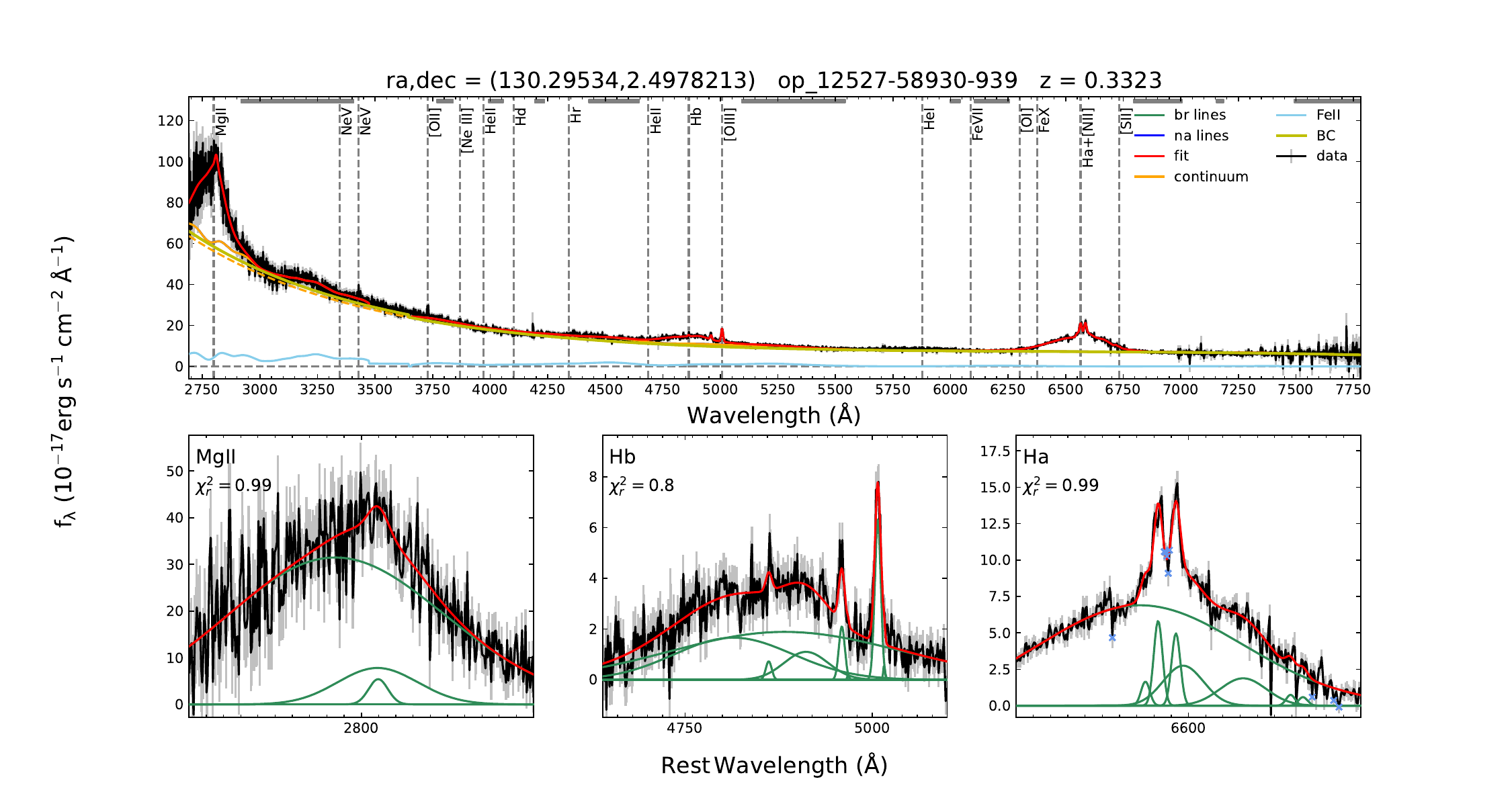}
\par\end{centering}
\caption{\label{fig:main-extra1}As in \cref{fig:optspec}, but for two OMBH quasars from the eFEDS-main sample (top: eFEDS-main ID 583, bottom: eFEDS-main ID 33). Both show very massive black holes through broad H$\beta$ lines.}
\end{figure*}
\begin{figure*}
\begin{centering}
\includegraphics[width=0.85\textwidth,trim={0 1.8cm 3cm 1.5cm},clip]{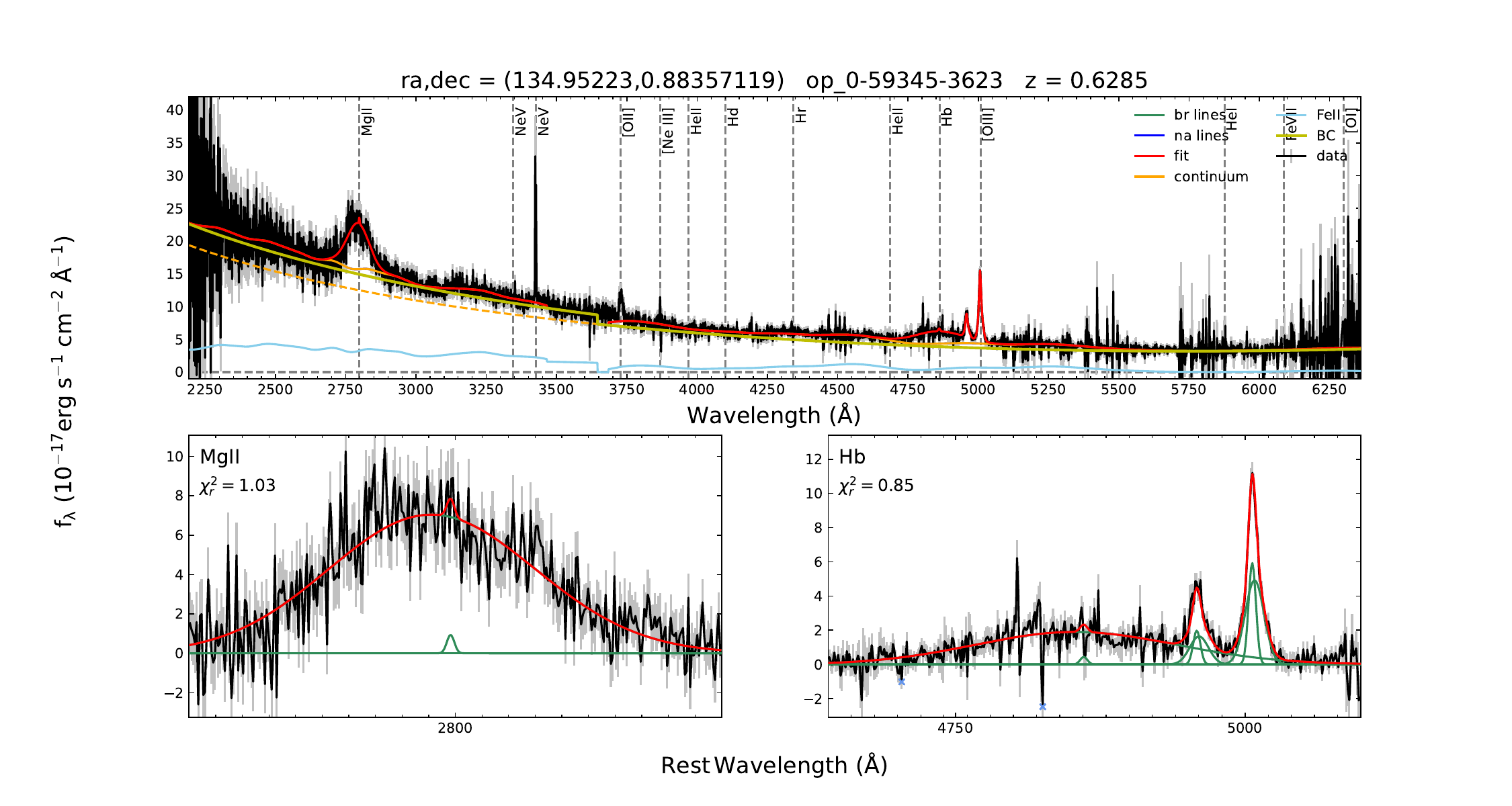}
\includegraphics[width=0.85\textwidth,trim={0 1.8cm 3cm 1.5cm},clip]{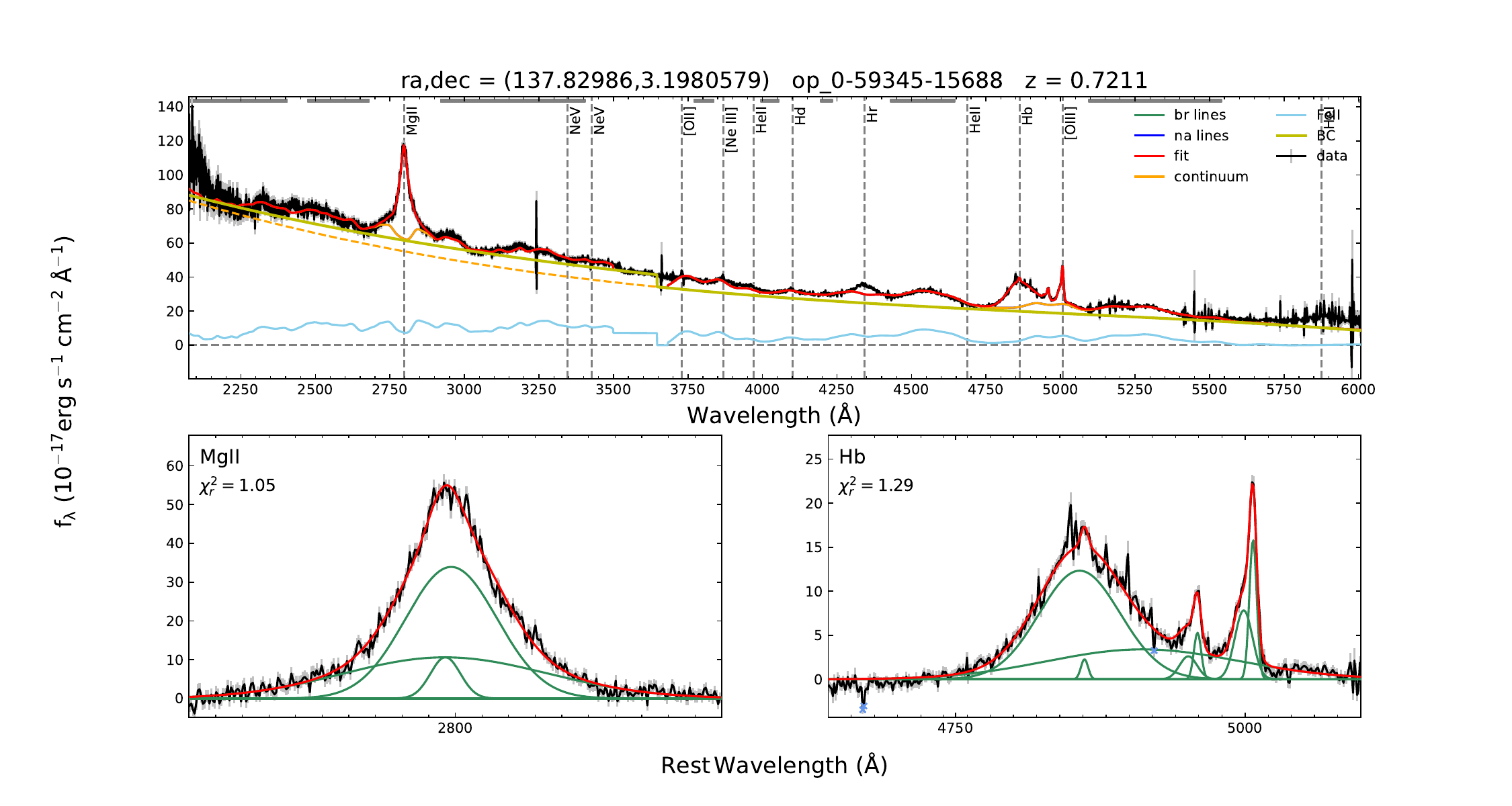}
\includegraphics[width=0.85\textwidth,trim={0 1.8cm 3cm 1.5cm},clip]{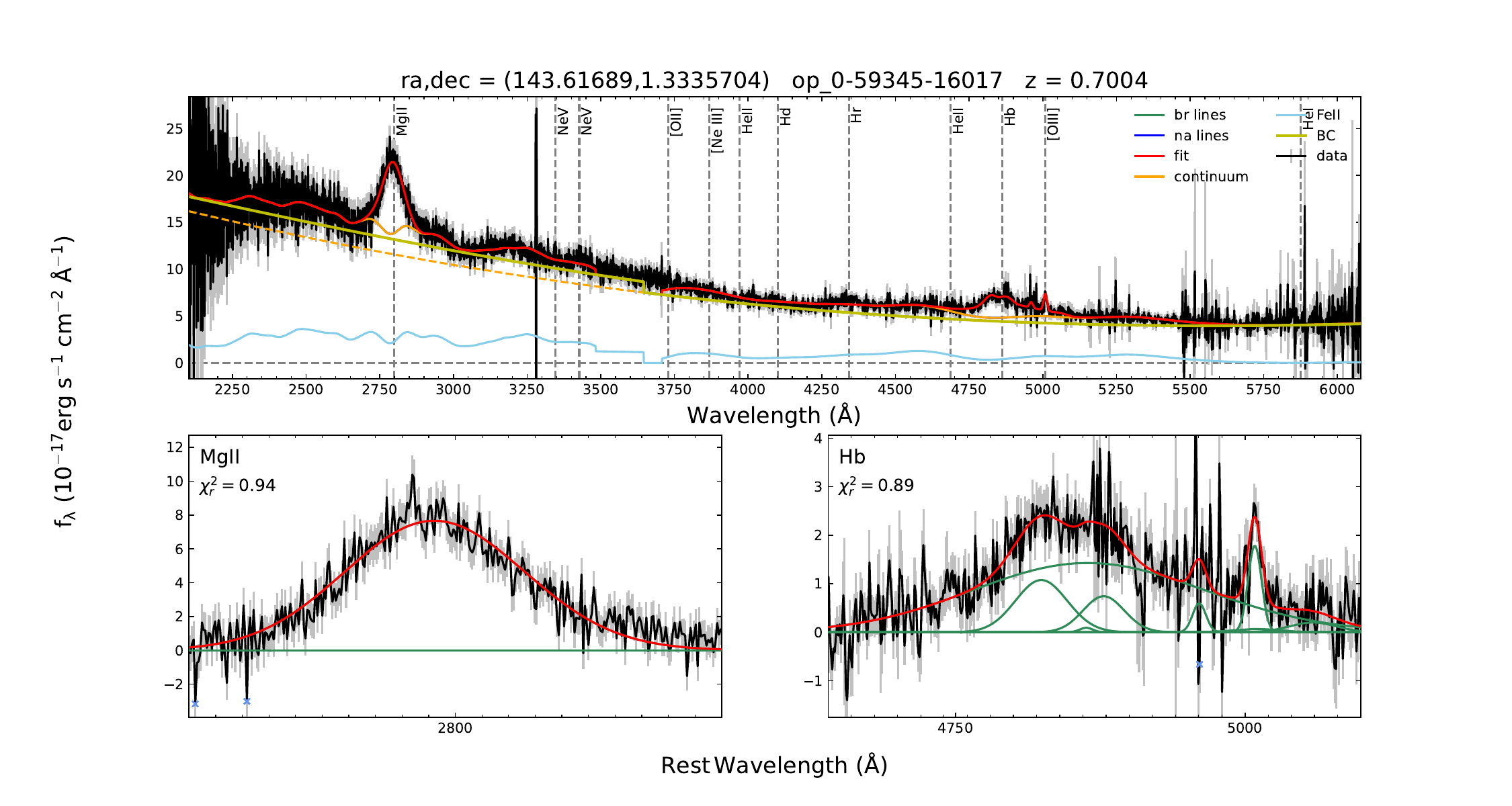}
\par\end{centering}
\caption{\label{fig:main-extra2}Optical spectra of three OMBHs in the eFEDS main catalog (top: eFEDS-main ID 3276, middle: eFEDS-main ID 689, bottom: eFEDS-main ID 553). All show very massive black holes through broad H$\beta$ lines.}
\end{figure*}

\begin{figure*}
\begin{centering}
\includegraphics[width=0.85\textwidth]{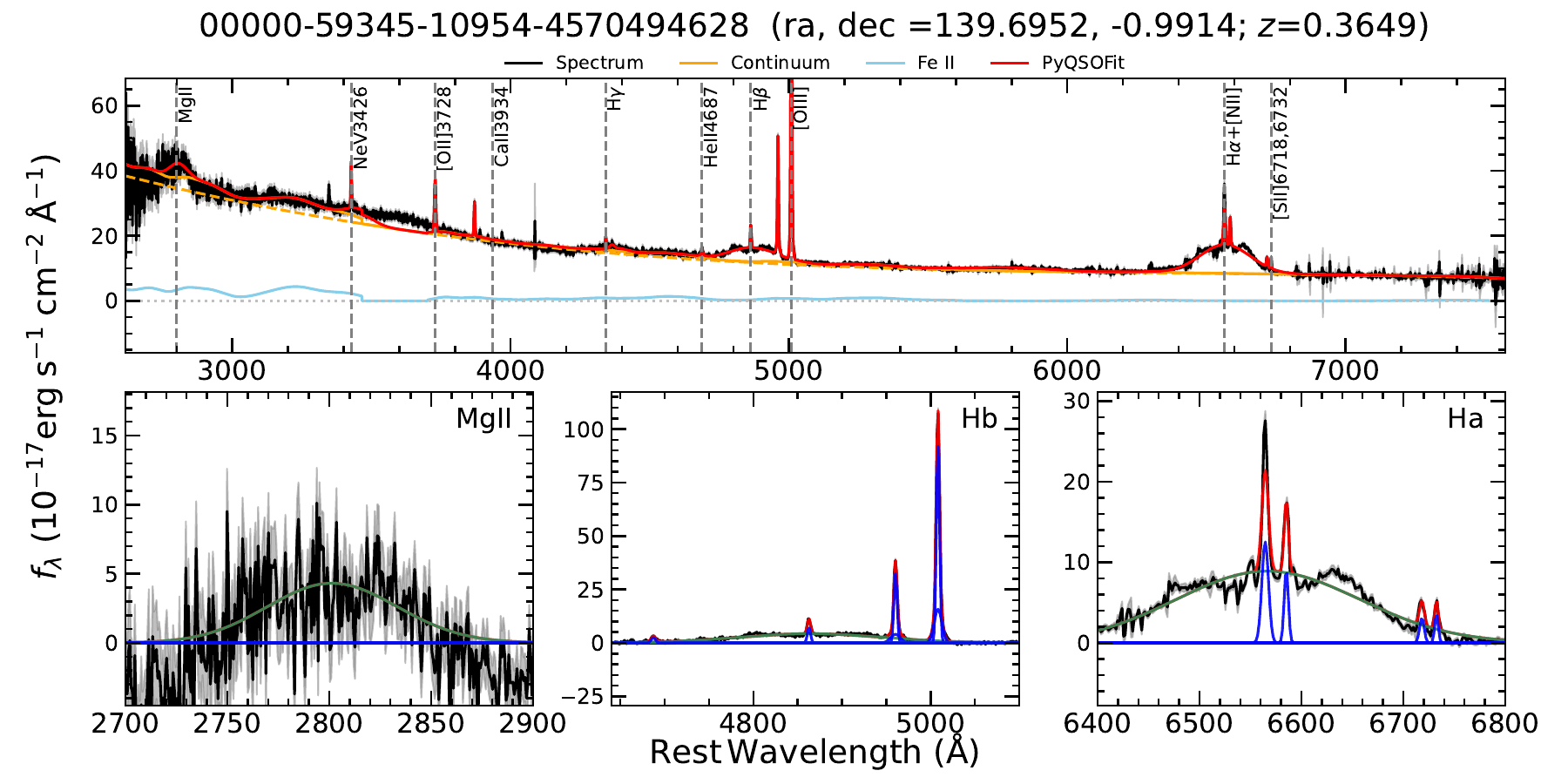}
\par\end{centering}
\begin{centering}
\includegraphics[width=0.85\textwidth]{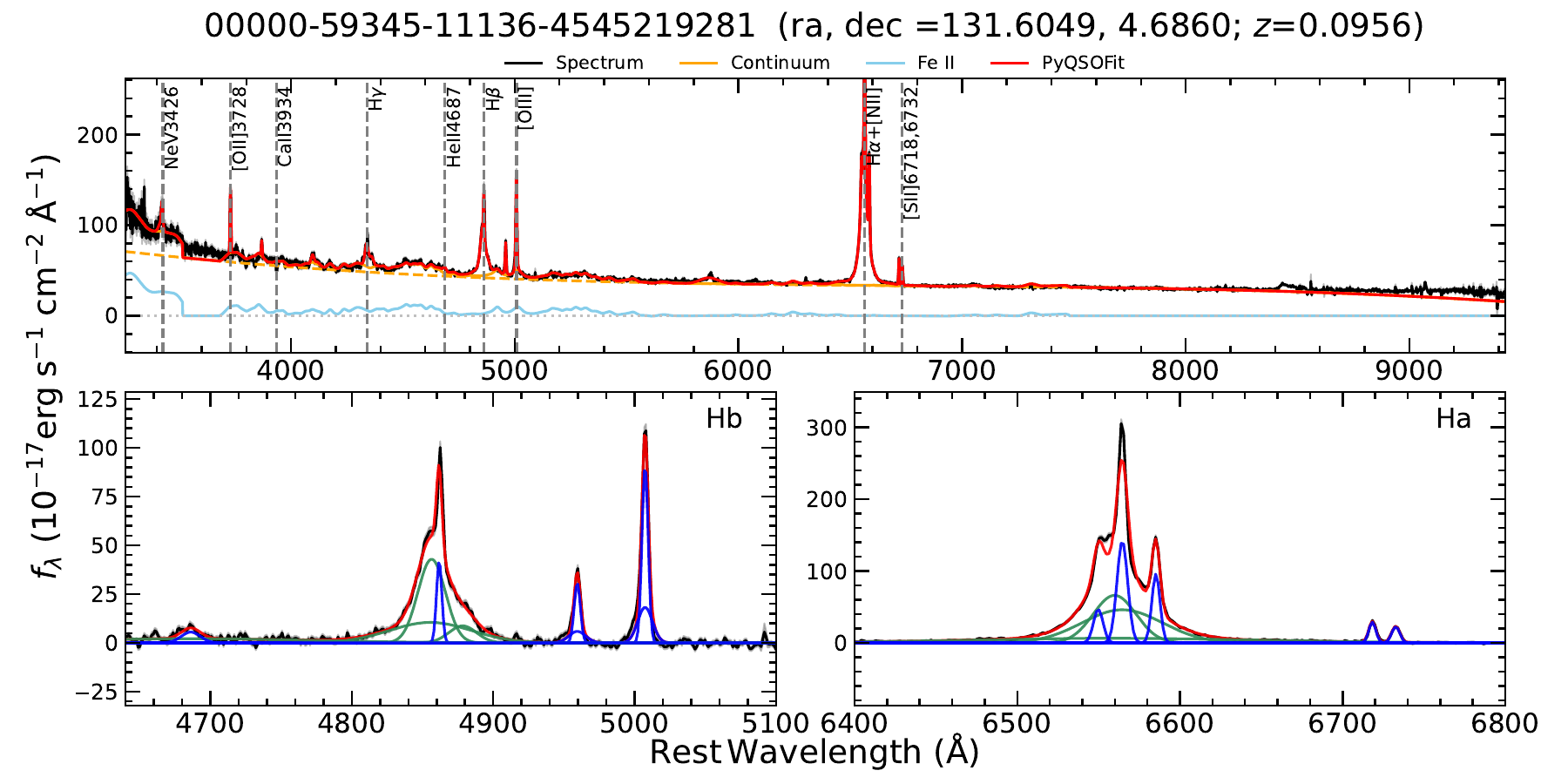}
\par\end{centering}
\begin{centering}
\includegraphics[width=0.85\textwidth]{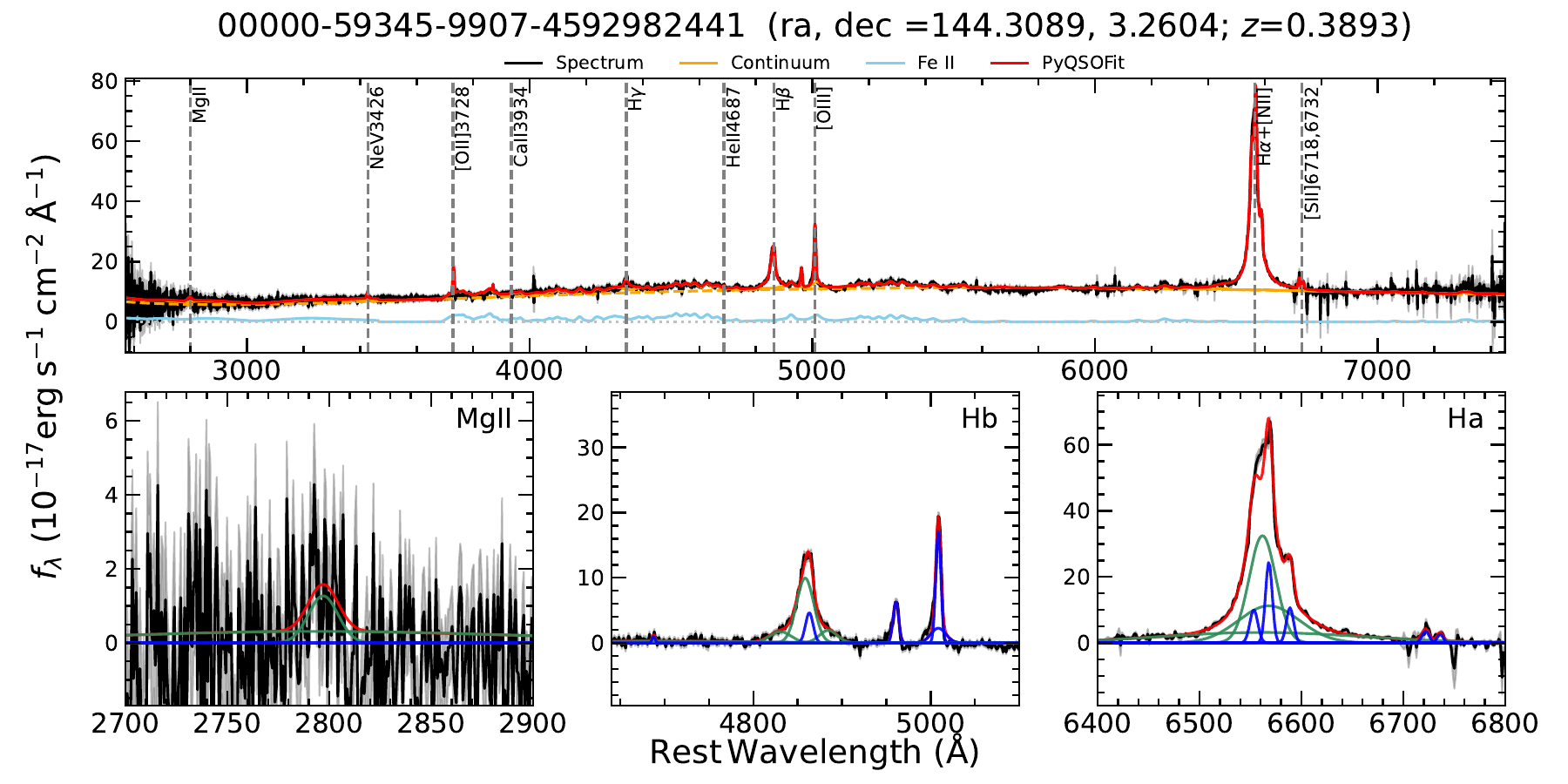}
\par\end{centering}
\caption{\label{fig:hard-extra1}Optical spectral fits (as in \cref{fig:optspec}) of eFEDS-hard ID 45 (top panel), the source with the fourth-largest
$M_{\bullet}/M_{\star}$ ratio in that sample, and two under-massive AGN, eFEDS-hard IDs 54 (middle panel) and 39 (bottom panel).}
\end{figure*}

Black hole masses are estimated from optical spectra. The SDSS spectra
are fit with PyQSOFit \citep{Guo2018} as described in \cite{Wu2022}.
The $H\beta$ FWHM and $5100\text{\AA}$ monochromatic luminosities
are converted into black hole masses with coefficients from \cite{Vestergaard2006}.
Following \cite{Wu2022}, we discard unreliable fits by requiring a
line flux signal-to-noise ratio (SNR) of larger than 2, a line luminosity
within the $L=10^{38-48}\mathrm{erg/s}$ range and that at least half of the pixels in each line complex remain after masking telluric-affected wavelengths.
Fig~\ref{fig:optspec} shows three example spectra of very massive black holes from the eFEDS-hard sample, while Fig~\ref{fig:hard-extra1} shows examples of lower mass black holes. Fig~\ref{fig:main-extra1} and \ref{fig:main-extra2} show example spectra from the eFEDS-main sample, which were treated in the same way (see Aydar et al., submitted). In these figures, the first step is continuum modelling (upper inset), which as described in \cite{Wu2022} combines a power law, polynomial components and a Fe II template (1000-3500$\AA$ and 3685-7484$\AA$). No host galaxy component is included, because in these quasar-dominated spectra this could cause systematic errors. After continuum model subtraction, each line complex (bottom insets) is modeled with up to three broad components (FWHM$>$1400\,km/s), a rest-frame narrow component (blue, FWHM$<$1400\,km/s), and additional contaminating lines \citep[see][]{Wu2022}. Narrow lines are at the same systemic velocity, for example in the H$\beta$ window, the velocity offset of the OIII line centers is tied to the velocity offset of the narrow H$\beta$ line component.
The broad components are then combined to measure the total full width at half maximum (FWHM). We visually
inspected the spectral fits for the modeling of the continuum around the lines and non-Gaussian line shapes. A few cases with poor or unphysically complex fits to the host continuum were discarded \citep{Nandra2025}, as in these cases the spectral line parameters are expected to be unreliable.
Mass estimates with statistical uncertainties smaller than 0.3~dex are retained.

Our single-epoch mass estimates have a systematic uncertainty of 0.45\,dex, as derived from reverberation mapping campaigns \citep[e.g.,][]{Shen2024}. 
This uncertainty stems in part from limited knowledge of the broad line region geometry and the fraction of the line emission attributable to virialized motion. This systematic affects virtually all studies of $M_\star-M_\bullet$ at substantial redshifts. Additional to that, JWST studies extrapolate the $z<2.5$ reverberation mapping campaign calibrations to $z>6$, well beyond the original luminosity and mass range, whereas this work lies mostly within the calibration regime. \Cref{sec:results:rm} below shows a comparison to direct black hole measurements efforts. The line asymmetries we find are largely within the distribution of the line shapes in reverberation mapping campaigns. However, in part due to the high signal-to-noise of our spectra, we find evidence for boxy and double-peaked broad lines in some of our sources. 
In the local Universe, double-peaked broad lines appear more commonly in massive galaxies \citep{Ward2025}, and thus are expected to be more common in the central engines of massive black holes. Whether the single-epoch mass estimate uncertainty is a function of the broad line profile is still unclear, in part because reverberation mapping samples are too small to robustly test for such an effect. The effect of the systematic uncertainty of single-epoch black hole mass estimates is addressed below with simulations.

We focus on $H\beta$ masses
as they have the highest reliability. 
For source IDs 32 and 133, the mass inferred from Mg II agrees closely with that from $H\beta$; for the others it is within the expected scatter (e.g., for IDs 1136 and 45, the Mg II masses are lower by a factor of 2.5).
The black hole masses are in
the range of $M_{\bullet}\sim10^{8-9.5}M_{\odot}$ and the Eddington
ratios are between $1\%$ and $100\%$ \citep{Nandra2025}.

\subsection{Total stellar mass estimate}\label{sec:method:stellarmass}

Measuring host stellar masses is challenging in systems with rapid
black hole growth. The tracer of the stellar mass, the host galaxy
spectral energy distribution (SED), is entangled with the broad SED
of the luminous AGN. To overcome biases in $M_{\star}$ estimates, \citet[][B24 hereafter]{Buchner2024} proposed a new Bayesian SED analysis method, GRAHSP (Genuine Retrieval of the AGN Host Stellar Population).
The approach uses a flexible AGN model. This avoids a failure mode of overly rigid AGN that cannot fully match the data, leading the galaxy template to compensate for the remaining mismatches and thereby bias the inferred stellar mass to high values in luminous AGN.
In this work, we follow the GRAHSP priors for AGN, galaxy, and attenuation parameters as described in B24 for their eFEDS analyses.
Briefly, the galaxy component is modeled
as a standard stellar population model \citep[using][]{Bruzual2003,Chabrier2003}
of solar metallicity with a $\tau$-delayed, exponentially declining star-formation history, attenuated by Small Magellanic Cloud-like dust \citep{Prevot1984}, with energy-conserving infrared re-emission \citep{Dale2014}. The AGN is modeled with a
power law for the big blue bump, augmented with line emission features (including Fe II), and a cold and hot dusty torus component. As this work focuses
on broad line AGN, we set the nuclear AGN attenuation (E(B-V)-AGN
parameter) to a low value (0.01) when the lower $2\sigma$ error bar
of the line width of the $H\beta$, $H\alpha$, MgII or $C\mathrm{IV}$
broad component lies above $1500\,\mathrm{km/s}$. In any case the
total attenuation parameter E(B-V) of the system is free to vary. We refer the reader to B24 for more details.

\begin{figure}
\centering

\begin{subfigure}{0.32\columnwidth}
\centering
\begin{tikzpicture}
\node[anchor=south west, inner sep=0] (img) 
    {\includegraphics[width=\linewidth]{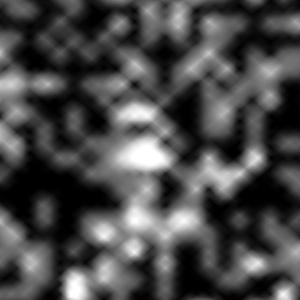}};
\begin{scope}[x={(img.south east)}, y={(img.north west)}]

\node[anchor=north west, text=white, font=\bfseries] at (0.02,0.98) {GALEX};

\draw[white, line width=2pt] (0.05,0.08) -- ++(0.033,0);
\node[anchor=south, text=white] at (0.0665,0.09) {$1''$};

\end{scope}
\end{tikzpicture}
\end{subfigure}
\hfill
\begin{subfigure}{0.32\columnwidth}
\centering
\begin{tikzpicture}
\node[anchor=south west, inner sep=0] (img) 
    {\includegraphics[width=\linewidth]{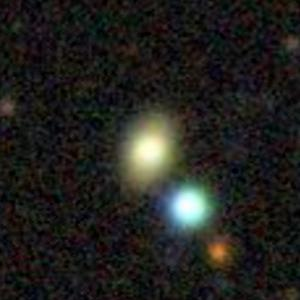}};
\begin{scope}[x={(img.south east)}, y={(img.north west)}]

\node[anchor=north west, text=white, font=\bfseries] at (0.02,0.98) {DECAM};

\draw[pink, line width=0.1pt] (0.5,0.5) circle (0.0167); % 0.5"
\draw[pink, line width=0.1pt] (0.5,0.5) circle (0.0250); % 0.75"
\draw[pink, line width=0.1pt] (0.5,0.5) circle (0.0333); % 1.0"
\draw[pink, line width=0.1pt] (0.5,0.5) circle (0.0500); % 1.5"

\draw[white, line width=2pt] (0.05,0.08) -- ++(0.033,0);
\node[anchor=south, text=white] at (0.0665,0.09) {$1''$};

\end{scope}
\end{tikzpicture}
\end{subfigure}
\hfill
\begin{subfigure}{0.32\columnwidth}
\centering
\begin{tikzpicture}
\node[anchor=south west, inner sep=0] (img) 
    {\includegraphics[width=\linewidth]{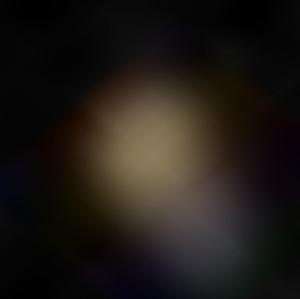}};
\begin{scope}[x={(img.south east)}, y={(img.north west)}]

\node[anchor=north west, text=white, font=\bfseries] at (0.02,0.98) {WISE};

\draw[white, line width=2pt] (0.05,0.08) -- ++(0.033,0);
\node[anchor=south, text=white] at (0.0665,0.09) {$1''$};

\end{scope}
\end{tikzpicture}
\end{subfigure}

\vspace{0.5em}

\begin{subfigure}{\columnwidth}
\centering
\includegraphics[width=\textwidth]{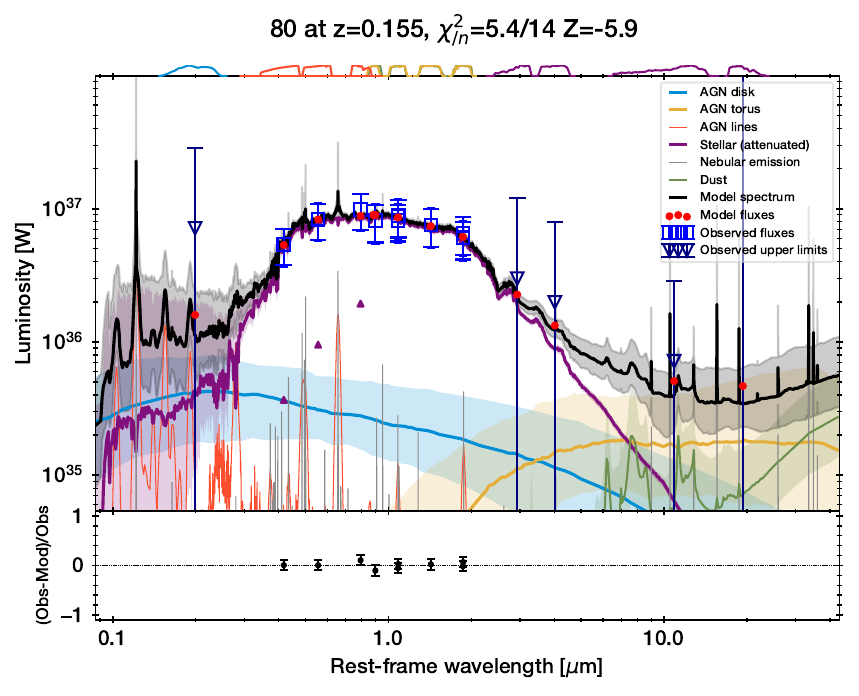}
\end{subfigure}

\caption{\label{EFLL-example}Main panel: Example GRAHSP SED fit with EFLL constraints on the host galaxy shown as purple up-wards pointing triangles.
Top row: GALEX (UV), DECam (optical), and WISE (IR) images, each with a $30''$ field of view. The UV images has low significance, giving a upper limit in the SED data, the WISE image is blended, giving upper limits on the total, leaving only the optical and near-infrared measurements.
The optical panel includes concentric apertures of 0.5, 0.75, 1.0, and $1.5''$ radius, the last is where extended flux beyond point source expectations was measured, which are constraints in the fit on the galaxy component (purple curve) alone.}
\end{figure}
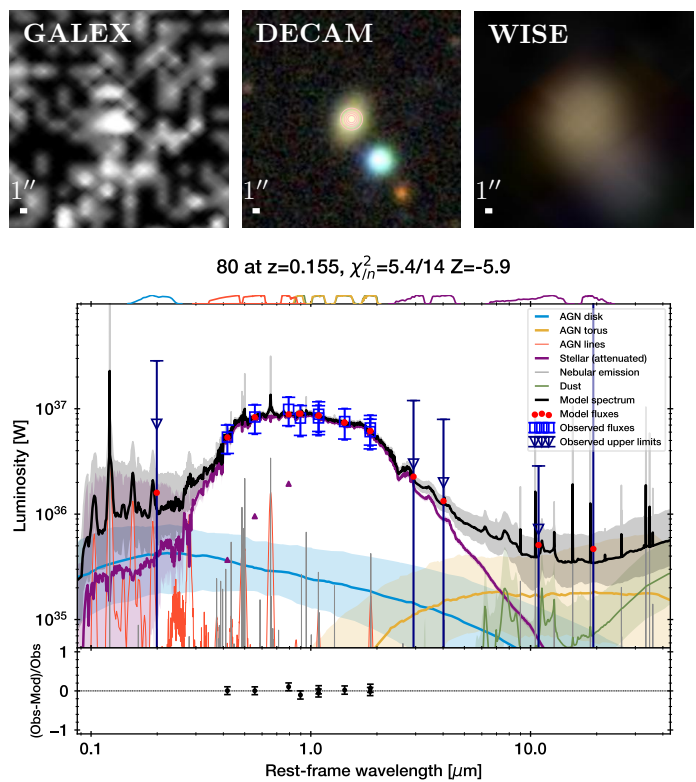

We enhance the GRAHSP SED fitting with a new morphology-independent
spatial decomposition technique which we call extended flux lower
limits (EFLL). For each AGN, we compare LS10 optical photometry fluxes extracted
in apertures\footnote{The centroid positions are from LS10 source detection, which performs multi-source source position and radial profile fitting on sky patches, see \cite{Dey2019}.} of radii of 0.5, 0.75, 1, 1.5, 2, 3.5 and 5 arcsec against
the expectation from a point sources matched in normalisation in the
0.5 arcsec annulus. We assume a Gaussian PSF model with the nominal
seeing for the relevant source and band. In our tests, this agrees well with the PSF observed in stars near to the AGN, and we get consistent results with either method. After subtraction, from the
observed aperture photometry in the 5 arcsec bin, we obtain an estimate
for the extended flux. We propagate the Gaussian uncertainties arising when matching
the central flux and from the 5 arcsec aperture flux. Finally, we use the mean $\mu_{b,\mathrm{EFLL}}$
and uncertainty $\sigma_{b,\mathrm{EFLL}}$ of the residual (extended) flux as a lower limit during SED fitting on the galaxy component alone,
with an additional upper limit likelihood term. GRAHSP can ingest
asymmetric Gaussian priors on any model parameter and model flux predictions
through the provided input file. For example, the column prior\_GALflux\_decam\_r
(and \_errlo/\_errhi) sets the mean $\mu$ (and lower/upper standard
deviation $\sigma_{r,+}$/$\sigma_{r,-}$) for the galaxy flux alone
in the decam r band. If $\sigma_{r,+}>0$ and $\sigma_{r,-}>0$, this
modifies the likelihood $L$ with the additional term:

\[
\ln L'=\ln L-\frac{1}{2}\begin{cases}
\left(\mathrm{galflux}_{\mathrm{r}}-\mu_{r}\right)^{2}/\sigma_{r,+}^{2} & \text{if }\mathrm{galflux}_{\mathrm{r}}\geq\mu_{r}\\
\left(\mathrm{galflux}_{r}-\mu_{r}\right)^{2}/\sigma_{r,-}^{2} & \text{if }\mathrm{galflux}_{\mathrm{r}}<\mu_{r}
\end{cases}.
\]
Here $\mathrm{galflux}_{\mathrm{r}}$ is the galaxy SED model flux
in the r band, and we set $\mu_{r}=\mu_{r,\mathrm{EFLL}}$, $\sigma_{r,-}=\sigma_{r,\mathrm{EFLL}}$.
Thus, during SED fitting, this gives low probability (highly negative
log-likelihood, high chi-square values) for fit proposals that lack
sufficient host galaxy flux in the band. On the other hand $\sigma_{r,+}$
is set to an arbitrary large number to make the likelihood term vanish
when the lower limit is fulfilled ($\mathrm{galflux}_{\mathrm{r}}\geq\mu_{r}$).
We repeat the validation test of section 7.3 of the GRAHSP paper and
compare the stellar masses estimated with the EFLL to masses from
spatially decomposed estimates from higher resolution
HyperSupremeCam data by \cite{Li2024}, and find consistency.
For about half of that validation sample, uncertainties are substantially reduced, often yielding informative stellar mass upper limits.

\begin{figure}
\centering
\includegraphics[width=\columnwidth,trim={8px 8px 10px 0},clip]{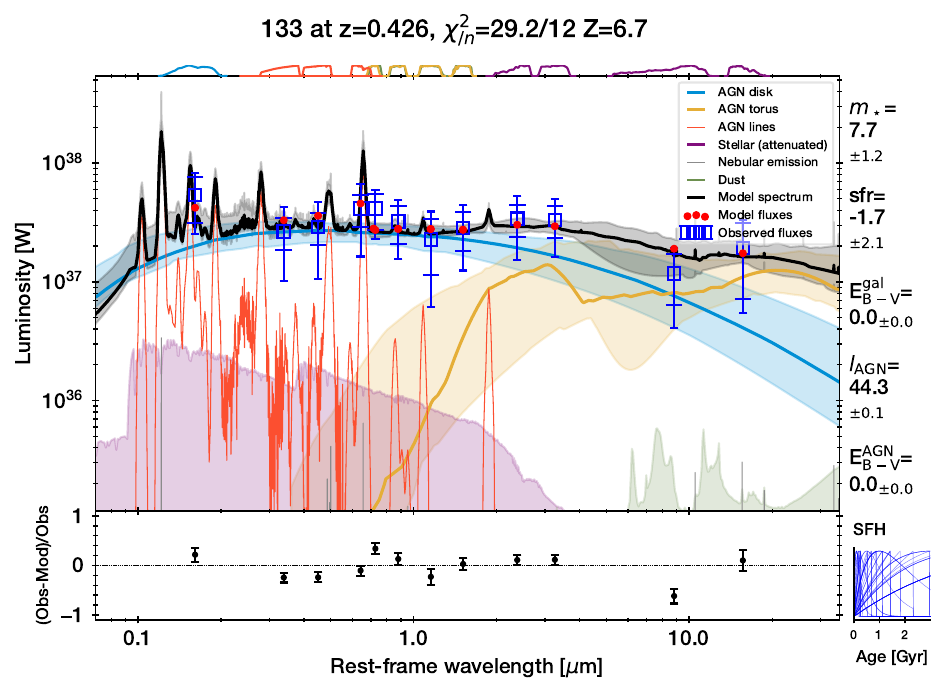}
\caption{\label{fig:SED}SED fits of ID 133, one of the three OMBH quasars from the eFEDS-hard sample. The parameters on the right are stellar mass in $M_\odot$, star formation rate in $M_\odot$/yr, attenuation and $5100\text{\protect\AA}$ AGN luminosity in erg/s (lower case labels indicate log units). The curves show best fit and $2\sigma$ uncertainties for each model component. The stellar mass posterior (violet band in the main panel shows the galaxy component) is wide and essentially an upper limit. All three sources are AGN-dominated, in agreement with the optical spectra in Figure\,\ref{fig:optspec}. On top of the panel, photometry filter transmission are drawn. The lower panel shows fit residuals. The bottom right inset panel shows posterior samples of the star formation histories, which are unconstrained.}
\end{figure}
\begin{figure}
\centering
\includegraphics[width=\columnwidth,trim={8px 8px 10px 0},clip]{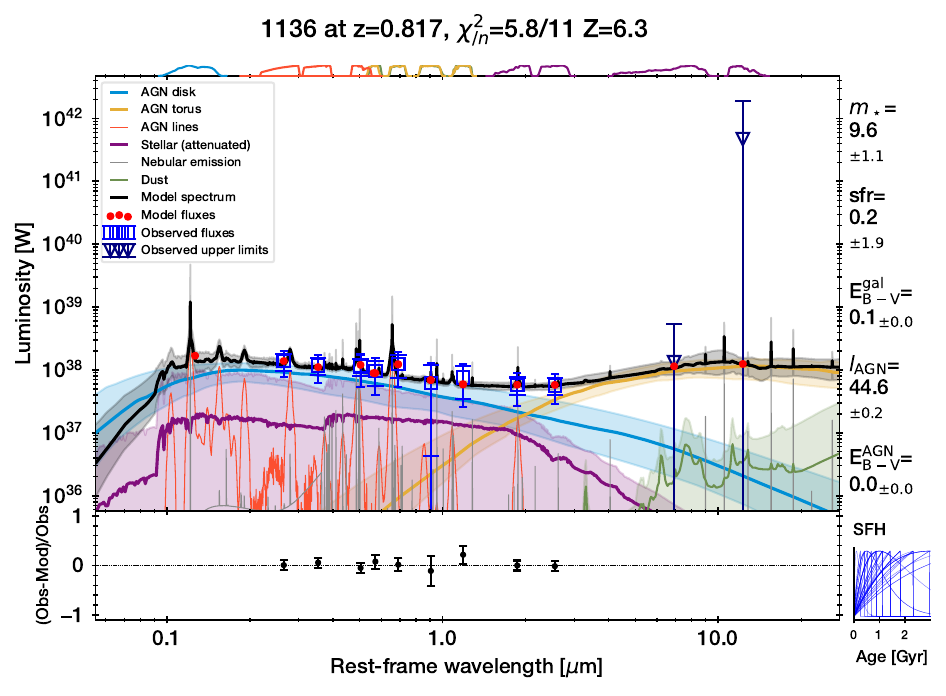}\\
\includegraphics[width=\columnwidth,trim={8px 8px 10px 0},clip]{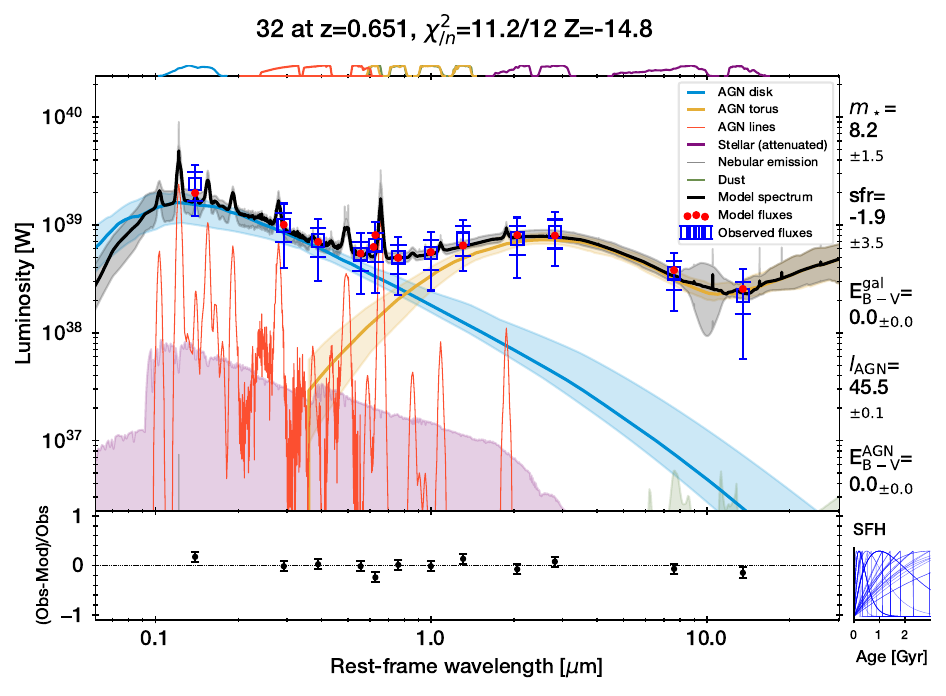}
\caption{\label{fig:SEDs}Same as Figure~\ref{fig:SED}, but for ID 1136 (top panel) and 32 (bottom panel), two of the three OMBH quasars from the eFEDS-hard sample. 
The stellar mass posterior (violet band in the main panel shows the galaxy component) is wide and essentially an upper limit. Both sources are AGN-dominated, in agreement with the optical spectra in Figure\,\ref{fig:optspec}.}
\end{figure}

For the parent sample considered in this work, we use RainbowLasso (B24)
to extract 5 arcsec diameter aperture-matched photometry from various surveys. This includes GALEX near-UV
\citep[NUV][]{Bianchi2017Galex}, optical from DESI Imaging Legacy Survey DR10 \citep[LS10 hereafter, based on DECam optical imaging and WISE IR;][]{Dey2019,Zenteno2025},
and near-infrared data from Vista Hemisphere Survey \citep{McMahon2013} and
UKIDSS \citep{Lawrence2007}. 
We incorporate EFLL constraints only in bands that meet RainbowLasso quality criteria and for which extended flux is detected ($\mu_{b,\mathrm{EFLL}}>0$). The hard X-ray band luminosity
is used as a prior on $5100\text{\AA}$ luminosity as described in
B24. Fitting the data with GRAHSP yields a host galaxy stellar mass
$M_{\star}$ estimate. GRAHSP reports $2\sigma$ uncertainties.

Figure~\ref{EFLL-example} illustrates the EFLL technique. The optical and near-infrared fluxes indicate a flat spectrum, which is degenerate between galaxy and AGN light. The UV data (top left panel in Figure~\ref{EFLL-example}) is of low significance, giving only a lower limit. The X-ray luminosity is tightly constrained and could inform the optical AGN template normalisation, however due to X-ray to optical conversion uncertainties this does not constrain our quasar-dominated optical photometry further, and thus does not give additional constraints the host galaxy. The infrared photometry is blended with a nearby source, and therefore deemed untrustworthy. Instead, RainbowLasso derives a conservative upper limit based on the sum of all components. This leaves a relatively short wavelength window of measurements available for the SED decomposition, and therefore a good case where spatial information can be helpful. The EFLL constraints are shown as purple upward-pointing triangles in Figure~\ref{EFLL-example}. In the z-band, the flux lower limit is closest to the total flux measurement, and thus most constraining. Proposed SED solutions where the purple curve falls below the purple up-ward pointing triangle are rejected during sampling. The posterior distribution of the galaxy component (purple shaded band) are small, although in this case, the Balmer break from g to r is also suggestive of a galaxy-dominated spectrum.

\subsection{Lauer bias simulations}\label{sec:method:lauer}

Flux-limited AGN samples of black holes incur selection effects \citep[e.g.][]{Lauer2007,Merloni2010}. We create a simulation to demonstrate the effects on our sample. In this section we only use the hard X-ray sample ($>2\,\mathrm{keV}$), where the selection function modeling is less affected by obscuration and additional soft X-ray emission components \citep{Waddell2024SX}.

Starting with a galaxy population model, we assign black hole masses following \cite{Kormendy2013} for quiescent galaxies and \cite{Greene2019} for star-forming ones, factoring in relation scatter. For the evolving galaxy population model, we adopt Schechter mass functions and quiescent fractions from \cite{Ilbert2013}.
Next, we trigger some galaxies to be AGN. 
We use a power-law Eddington ratio distribution \citep[c.f.,][]{Kelly2013,Georgakakis2017} 
with the normalization adjusted to match our redshift distribution. We converted AGN properties (luminosity, redshift, obscuration) to observed 2.3-5 keV fluxes assuming a log-uniform column density distribution, a constant X-ray bolometric correction of 25 for simplicity and the \texttt{UXClumpy} obscurer model \citep{Buchner2019b}.
The eFEDS sensitive area selection function \citep{Brunner2021} was applied to mimic observed samples.
Our simulated sample agrees with the observed sample in its luminosity and redshift distribution.
The effect of single-epoch black hole mass estimates is modeled by adding 0.45 dex in scatter to $M_\bullet$.

\section{Results}\label{sec:results}

\subsection{Identification of overmassive black holes}
\begin{figure*}
    \centering
    \includegraphics[width=0.8\linewidth]{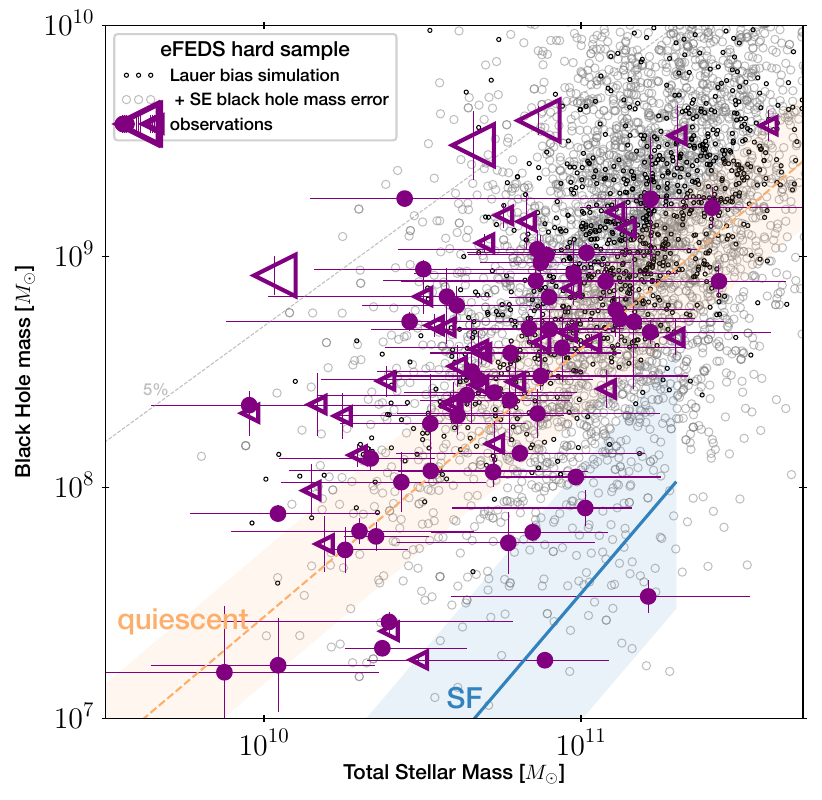}
    \caption{Our sample (purple) in comparison to scaling relations.
    Starting from star-forming and quiescent galaxy populations following local mass scaling relations, and an Eddington ratio distribution, the black circles indicate the AGN distribution expected to be detected by eROSITA. The 4000 gray circles indicate the distribution after applying the single-epoch mass estimation systematic scatter. Our observed sample (purple) is consistent with the expected distribution. Low-mass galaxies ($M_\star<10^{10}\,M_\odot$) with high black hole masses ($M_{\bullet}\sim10^9\,M_\odot$) at the top left are most difficult for the simulation to reproduce. The gray dashed indicates the threshold for our definition of OMBHs, $M_{\bullet}/M_\star>5\%$.}
    \label{fig:lauer}
\end{figure*}

In the top-left corner of Figure~\ref{fig:lauer}, we identify three systems 
that are about ten times above the local scaling relation of \cite{Kormendy2013},
with eFEDS-hard catalog ID 133, 1136 and 32. Figures~\ref{fig:SED} and \ref{fig:SEDs} shows the SED fits. 
The stellar mass upper limit is reliable: Conservatively converting the total z- or W1-band flux using an old-galaxy template yields similar stellar mass upper limits. Thus, regardless of AGN modeling details, the stellar mass cannot be substantially higher.
Further below, we characterize the physical properties of these OMBHs in more details. However, first we present results indicating that these sources are expected and the selection is robust.

\subsection{Lauer bias simulations}\label{sec:resultslauer}

We simulated the expected sample based on scaling relations and the evolving galaxy population.
To validate the simulation, the stellar mass and black hole mass distribution of 
1000 simulated systems is shown as black circles in 
Figure~\ref{fig:lauer}. Quiescent and star-forming galaxies contribute roughly equally across the $M_{\star}$ range. At higher $M_{\star}$, the simulations show more $M_{\bullet}>10^{9.5}\,M_\odot$ than the observed sample. This arises from the assumed larger scatter in the star-forming scaling relation.

The gray circles in Figure~\ref{fig:lauer} show the effect of adding 0.45 dex single-epoch mass scatter to the 1000 simulated sources, which we oversample by a factor of 5. The median mass ratio is $M_{\bullet}/M_{\star}=1\%$. We then scale our simulated number of sources to 200 to gain estimates appropriate the eFEDS-hard sample.
In our realistic simulations, 17 out of 200 systems have $M_{\bullet}>5\%M_{\star}$. These are often at high stellar masses, because the simulations produce a stellar-mass distribution that is more right-skewed than the observed one. If we correct for this by reweighting the simulated sample by $1/M_\star$, the stellar mass distributions match, then 15 out of 200 systems are expected at observed $M_{\bullet}>5\%M_{\star}$. Half have an intrinsic (before black hole measurement uncertainties) of $M_{\bullet}>1.9\%M_{\star}$, i.e., four times above the ratio typical of quiescent galaxies. Twelve percent have an intrinsic ratio (without single-epoch mass systematic scatter) of $M_{\bullet}>5\%M_{\star}$, i.e., one out of eight are comparable in ratio to NGC1277 or NGC4486B.

Finally, we comment on the behaviour at the low stellar mass end.
As in other simulations \citep{Pacucci2023,Li2024a}, the black hole mass distribution is squeezed there by the intrinsic rarity of very massive black holes and the high luminosity threshold.
A further contributing factor is that in this mass regime, quiescent galaxies are rare, 
and the scaling relations of star-forming galaxies are steeper than unity there \citep{Greene2019} (see Fig.~\ref{fig:images}).
Comparing data to simulations, the most outstanding system is ID 133 at the top left of \Cref{fig:images}. 

\subsection{The sample lies within $M_\bullet$ calibration distribution}\label{sec:results:rm}

\begin{figure}
\begin{centering}
\includegraphics[width=\columnwidth]{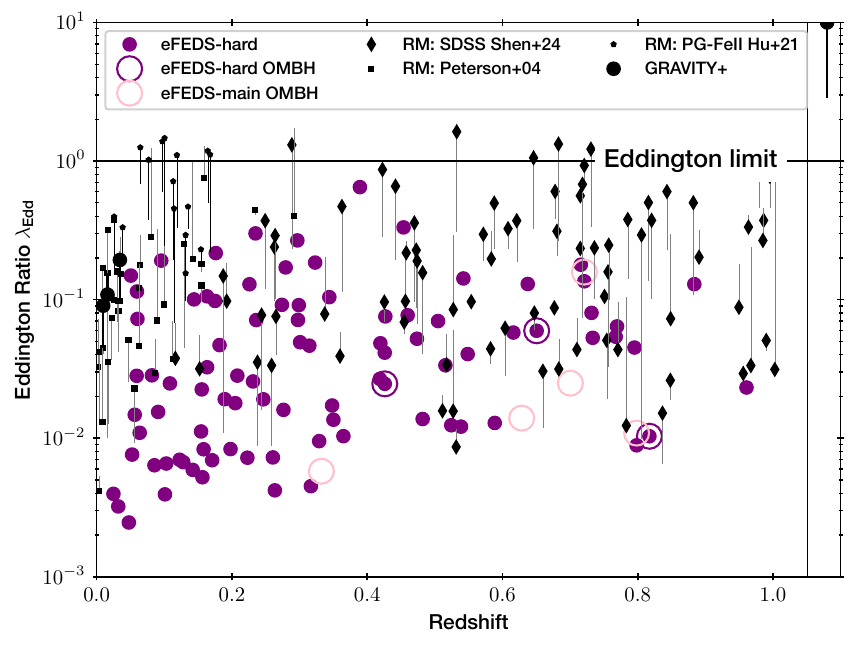}
\par\end{centering}
\caption{\protect\label{fig:lambdadist}Eddington ratio - redshift distribution.
The Eddington ratio is always computed from the luminosity at $5100\AA$, and black hole masses are all based on the $H\beta$ line.
Our sample (purple and pink symbols) uses single-epoch mass estimates, which have been calibrated from reverberation mapping campaign masses (black points \citep{peterson2004,Hu2021,Yue2024}). The thin lines connect the RM-derived mass (symbol) with the SE mass estimate.
Direct black hole mass measurements \citep{GRAV2024,GRAV2025Li} are shown in black circles, with the recent z=4 super-Eddington source at the top right \citep{GRAV2025}.
Our sample lies within the RM calibration range and below the Eddington limit.
The OMBHs (large circles) objects lie within the distribution of the parent eFEDS sample.
}
\end{figure}

This section highlights that for identifying OMBHs we operate within the distribution where the single-epoch (SE) mass estimates are calibrated and most reliable. Figure~\ref{fig:lambdadist} shows that the redshift distribution of the eFEDS sample with $H\beta$ mass estimates ranges from $z=0-1$, due to the SDSS spectrograph wavelength range. Black diamonds, pentagons and squares in Figure~\ref{fig:lambdadist} represent reverberation mapping (RM) campaigns \citep{peterson2004,Hu2021,Yue2024}, which cover the same redshift range. The y-axis shows the Eddington ratio computed for all points in this plot as $\lambda_\mathrm{Edd}=BC \times \lambda \sqrt{L_\lambda(5100\AA)} \mathrm{FWHM}(H\beta)^2$ with a bolometric correction of BC=9.26 \citep[following][]{Shen2011,Richards2006bolom}.
Each RM-mass symbol is connected to the corresponding single-epoch mass, indicating the typical scatter. As noted previously, the scatter increases toward higher $\lambda_\mathrm{Edd}$ \citep{peterson2004,Hu2021,Yue2024}

\begin{figure*}
\begin{centering}
\includegraphics[width=0.99\textwidth]{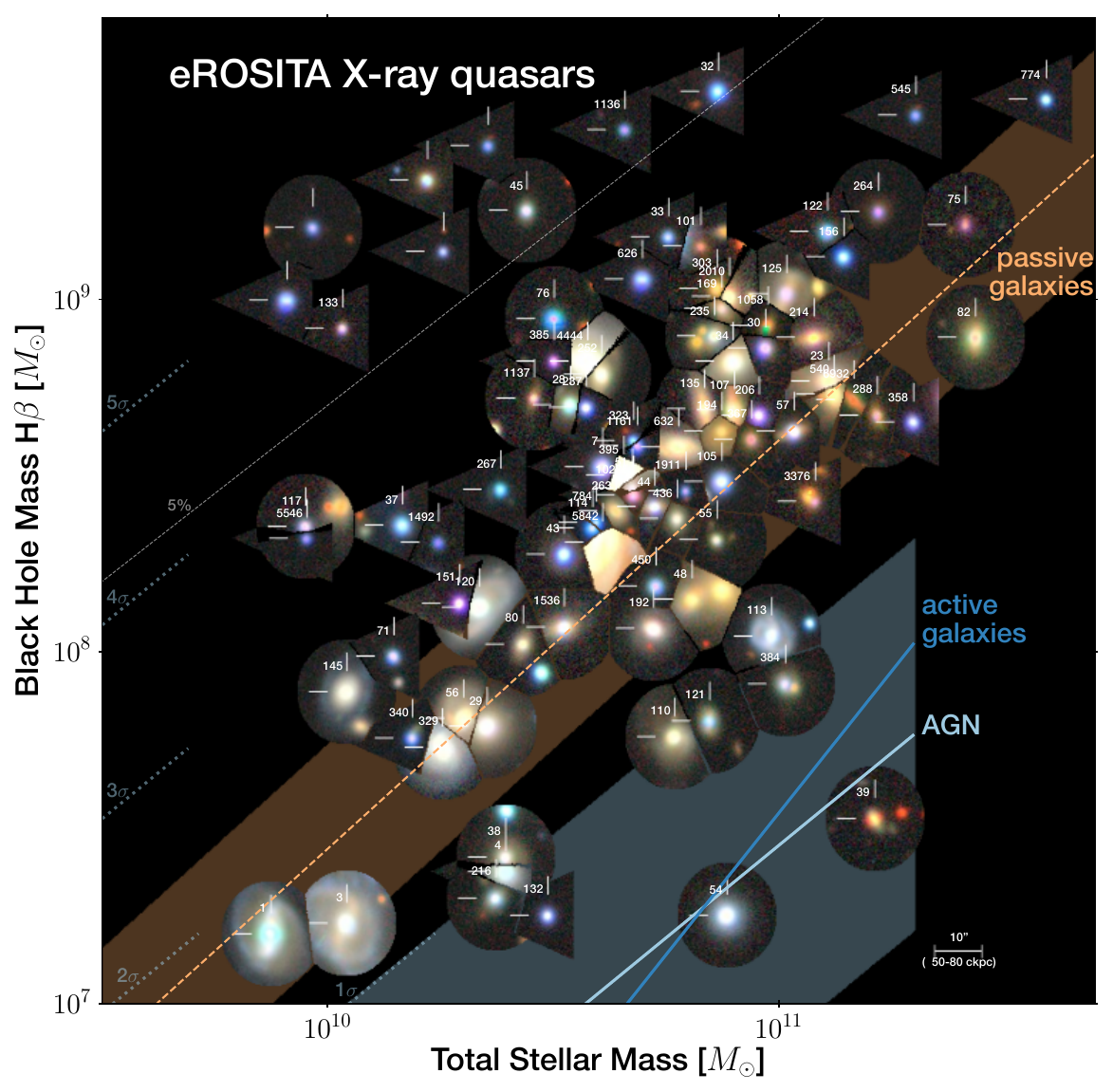}
\par\end{centering}
\caption{\protect\label{fig:images}Optical cutout (grz LS10 10$''$ radius) for each X-ray quasar
placed on the $M_{\bullet}-M_{\star}$ plane.
The orange dashed line shows the local galaxy scaling relations of elliptical galaxies and bulges \citep{Kormendy2013}.
For gas-rich systems the relation lies further to the bottom right, 
with star-forming galaxies (dark blue line) \citep{Greene2019} and local optical AGN \cite{Reines2015} (light blue line).
From the 1$\sigma$ scatter of the local AGN relation (light blue shaded region),
our overmassive systems ($M_\bullet/M_\star>5\%$, dotted line) on the top left lie four standard deviations ($4\sigma$) away.
Left-ward pointing triangles are upper limits in stellar mass. Sources with IDs are from the eFEDS-hard sample, the remaining five are from the eFEDS-main sample.
}
\end{figure*}

\begin{table*}[]
    \centering
    \caption{OMBHs ($M_\bullet/M_\star>5\%$) found in the eFEDS-hard survey (top) and eFEDS-main survey (bottom). The column give the optical coordinates, redshift, black hole mass, stellar mass and 0.5-2keV rest-frame AGN luminosity. %\textcolor{red}{TODO: add Eddington ratio to show that these are sub-Eddington.}
    }
    \label{tab:sources}
    
\begin{tabular}{r p{1.2cm} p{1.1cm} p{0.5cm} p{0.4cm} ll}
Hard Sample ID &RA&Dec&z&$M_{\bullet}$&$M_\star$&$L_X$\\
\hline
\hline
 &deg&deg& &$M_\odot$&$M_\odot$&erg/s\\
 & J2000 && &log&log&log\\
\hline
\hline
133&143.7766&3.7543&0.43&8.9&$<10.0$&$44.3_{44.2}^{44.4}$\\
1136&129.7403&4.2407&0.82&9.5&$<10.7$&$44.4_{44.2}^{44.5}$\\
32&143.8258&2.0710&0.65&9.6&$<10.9$&$45.1_{45.1}^{45.1}$\\
45&139.6952&-0.9914&0.36&9.3&$10.4_{10.1}^{10.7}$&$44.1_{44.1}^{44.2}$\\
%33&136.1917&2.1453&0.80&9.2&$<10.8$&$45.2_{45.2}^{45.3}$\\
%5546&128.2232&1.2734&0.62&8.3&$<10.0$&$43.8_{43.6}^{44.0}$\\
%626&137.8299&3.1981&0.72&9.1&$<10.7$&$44.7_{44.6}^{44.7}$\\
%385&140.8790&-0.0453&0.48&8.8&$<10.5$&$44.1_{44.0}^{44.2}$\\
%101&128.5106&-1.8125&0.54&9.2&$<10.8$&$44.5_{44.4}^{44.5}$\\
%545&136.5168&1.1873&0.80&9.5&$<11.3$&$44.6_{44.5}^{44.7}$\\
%1&144.2544&1.0955&0.05&7.2&$9.9_{9.5}^{10.4}$&$43.1_{43.1}^{43.2}$\\
\hline
\hline
Main Sample ID &RA&Dec&z&$M_{\bullet}$&$M_\star$&$L_X$\\
\hline
\hline
33&130.2953&2.4978&0.33&9.3&$9.4_{8.3}^{10.2}$&$44.3_{44.3}^{44.3}$\\
689&137.8299&3.1981&0.72&9.0&$<9.9$&$44.2_{44.2}^{44.3}$\\
583&136.5168&1.1873&0.80&9.4&$8.9_{5.3}^{10.4}$&$44.5_{44.5}^{44.6}$\\
553&143.6169&1.3336&0.70&9.1&$<10.3$&$44.5_{44.4}^{44.5}$\\
3276&134.9522&0.8836&0.63&9.2&$10.0_{9.3}^{10.4}$&$43.8_{43.7}^{43.9}$\\
\end{tabular}

\end{table*}

Figure~\ref{fig:images} presents the optical images of our parent sample of \ero{} hard X-ray selected quasars in the eFEDS 140 deg$^2$ survey area \citep{Brunner2021,Salvato2021,Nandra2025}  positioned on the $M_{\bullet}-M_{\star}$ plane.
On this plot, the systems are centered at their black hole mass and stellar mass estimate.
In case of upper limits on the stellar masses, the cut-outs are left-ward pointing triangles. 
Sources with statistical errors larger than 1 dex in stellar mass or 0.3 dex in black hole mass are excluded.
We find a wide distribution, which is consistent with the population of 
star-forming and quiescent galaxies in combination with their respective mass 
scaling relations (blue and orange), when considering their intrinsic 
scatter \citep{Reines2015,Greene2019}, black hole mass measurement scatter and the Lauer bias. 
In the top left of Figure~\ref{fig:images} above the dashed line 
we identify three systems (ID: 133, 1136, 32; details in Table~\ref{tab:sources}) with 2 sigma lower limits 
on the mass ratio of $M_{\bullet}/M_{\star}>5\%$, which is ten times above 
the local quiescent galaxy scaling relation 
\citep{Kormendy2013,Greene2019}. 
One of the three cases, the source is not merely a blue point source, but shows either spatial extent, another shows reddened colors (see Appendix~\ref{sec:HSCimages} for radial profiles), indicative that the source is not entirely dominated by an unobscured accretion disk. 
Additionally, from the same survey area, but only detected in the soft \ero{} X-ray band, we have identified five additional OMBH quasars (details in  Table~\ref{tab:sources}). They are shown without IDs in the top left of Figure~\ref{fig:images}), also well above the quiescent scaling relation.  These five systems have similar properties to the three hard sample sources, with only one showing a potential nearby neighbor.
The comparison to star-forming galaxies \citep{Greene2019} is also interesting, because their cold gas may power luminous AGN somewhat more efficiently \citep{Bongiorno2012,Aird2019}. In that case, our eight systems lie more than 4$\sigma$ above the scaling relations of star-forming galaxies \citep{Reines2015}.
The systems are even more extreme in 
comparison to the steeper relation of local AGN 
\citep{Reines2015}, which features a similar scatter and already 
contains the single-epoch mass uncertainty. Our eight AGN are extreme 
systems, likely the tail of the galaxy population in black hole mass to stellar mass ratio at these redshifts.

\subsection{Comparison to literature samples}\label{sec:results:lit}

\begin{figure}
\centering
\includegraphics[width=0.99\columnwidth]{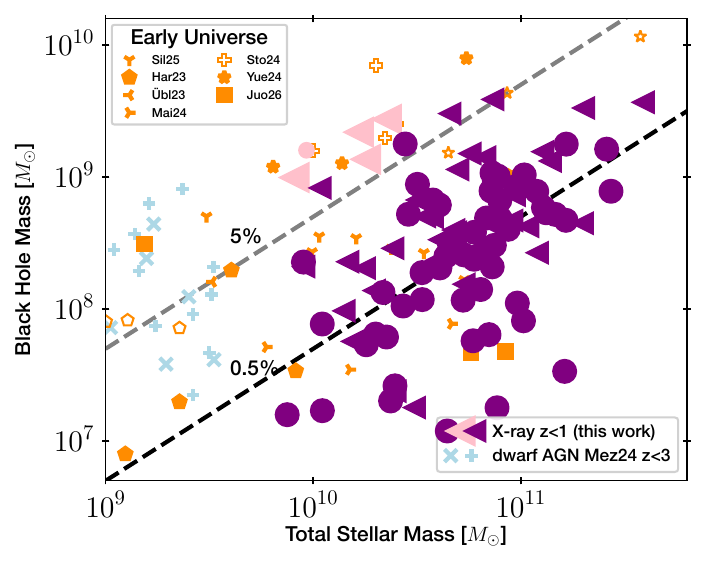}
\caption{\protect\label{fig:litmasseshighz}
Our data points compared to high-redshift OMBHs.
Our three (five) OMBH quasars from the eFEDS hard (main) X-ray survey are shown as large purple (pink) symbols.
Left-ward pointing triangles and empty symbols are upper limits in stellar mass ($2\sigma$ for ours).
Orange data points are from $z>5.5$ AGN from the SHELLQ sample \citep{Matsuoka2016,Silverman2025}
and JWST \citep{Ding2023,Harikane2023,Ubler2023,Maiolino2024,Stone2024,Yue2024,Juodzbalis2026}.
The distributions of the orange and purple/pink symbols overlap, both exceeding the 5\% threshold (gray dashed line).
At the very left, the type-1 AGN in dwarf
galaxies \citep{Mezcua2023,Mezcua2024} at $z<1$ ($z=1-3$) are shown as light blue crosses (pluses).
Left-pointing triangles are consistent dwarf galaxies as well.
}
\end{figure}

A closer comparison to other OMBHs is shown in 
Figure~\ref{fig:litmasseshighz}.
Our OMBHs (purple and pink triangles in Fig.~\ref{fig:litmasseshighz}) at $z<0.8$
cover the same mass ratios as those found in the first 
half billion years (orange points). Our distribution of measurements is 
limited to stellar masses of $M_{\star}>10^{10}M_{\odot}$,
however, the stellar mass upper limits are also consistent with dwarf galaxies.
Figure~\ref{fig:litmasseshighz} also shows dwarf-galaxy OMBH samples.

\begin{figure}
\centering
\includegraphics[width=0.99\columnwidth]{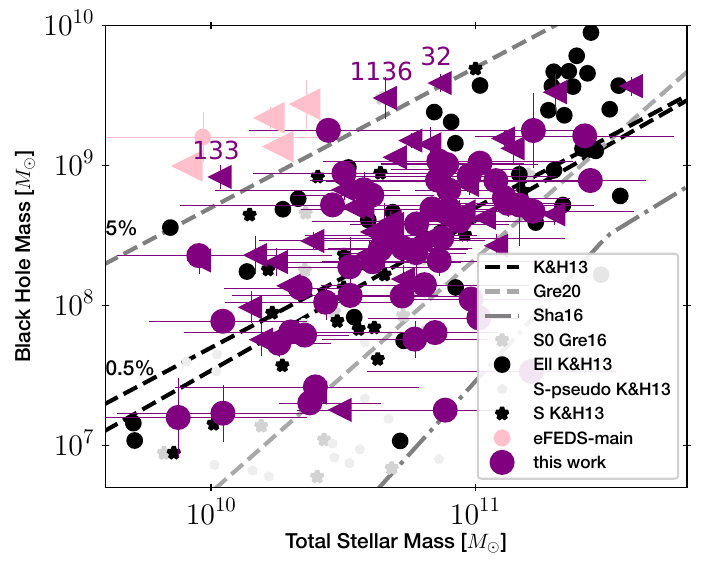}
\caption{\protect\label{fig:litmassesgal}
Our data points compared to non-AGN galaxies (symbols) and relations (lines) in the local Universe \citep{Kormendy2013,Reines2015,Saglia2016,Greene2016,Shankar2016}
for various morphological types. 
}
\end{figure}
\begin{figure}
\centering
\includegraphics[width=0.99\columnwidth]{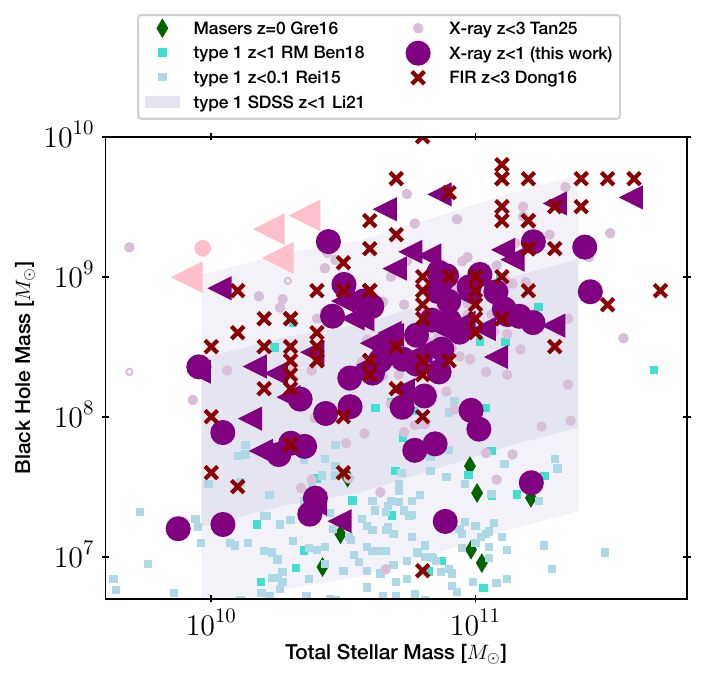}
\caption{\protect\label{fig:litmasses}
Our data points in comparison with other AGN samples. 
SDSS type-1 AGN at z=0.2-0.8 \citep[blue shading and cyan rectangles;][respectively]{Li2021a,Bentz2018} 
lie at similar locations as our sample,
while at $z<0.1$ SDSS type-1 AGN lie well below \citep[light blue squares;][]{Reines2015},
as do water megamasers \citep[green diamonds;][]{Greene2016}.
Herschel far-infrared selected luminous AGN at z=0-3 \citep{Dong2016}
are shown as red crosses. The X-ray type 1 sample at cosmic noon of \cite{Tanaka2025} is shown in purple circles.}
\end{figure}

The detection of black holes with $M_{\bullet}/M_{\star}\sim10\%$, far exceeding the average values of local scaling
relations, has prompted debates about extraordinary growth histories. 
Meanwhile, overmassive black holes found in dwarf galaxies are often considered
dormant remnants of black hole seeding \citep{Volonteri2010,Greene2019}. 
However, this scenario is less feasible for our OMBHs in 
massive, $z<1$ galaxies with $M_{\star}\geq10^{10}M_{\odot}$. These galaxies 
have undergone numerous galaxy-galaxy interactions \citep{Buchner2019b}, 
and the black hole mass has evolved substantially beyond the initial seed mass.
Therefore, it is more plausible that a different mechanism is responsible 
for the build-up of OMBHs. 
The presence of another mechanism, however, may erode support for the black hole seed interpretation for both nearby dwarf galaxies and distant galaxies observed by JWST.

Figure~\ref{fig:litmassesgal} compares our OMBHs to local non-AGN galaxies.
The distribution are of similar range, in $M_{\bullet}$, $M_{\star}$ and $M_{\bullet}/M_{\star}$.
The aforementioned local OMBH remnants NGC1277 and NGC4486B are visible near the 5\% dashed line.
Figure~\ref{fig:litmasses} compares our sample to other AGN samples.
Similar objects as those in our survey have been found previously. 
However, these used much wider area surveys, an important distinction discussed further below.

\begin{figure*}
    \centering
    \includegraphics[width=\linewidth]{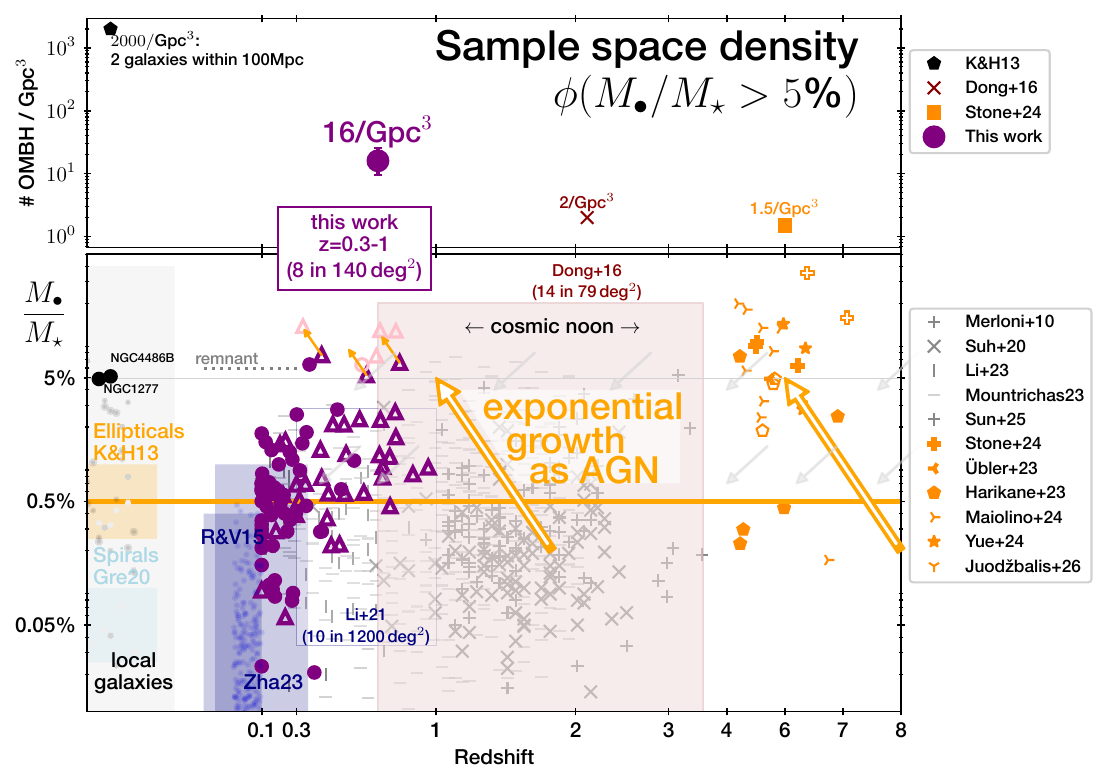}
    \caption{
    \textit{Top panel}: Detected space density of OMBHs over cosmic time. 
    We find a space density of ${16}_{-6.4}^{+9.7}/\mathrm{Gpc}^3$, which is higher than previous AGN surveys,
    but lower than the space density within 100 Mpc.
    \textit{Bottom panel}: Schematic sketch of growth of OMBHs over cosmic time. 
    Purple, pink and orange symbols are the same as in Fig.~\ref{fig:litmasses}. 
    This work (purple: eFEDS-hard, pink: eFEDS-main; circles: measurements, triangles: lower limits) finds OMBHs 
    after cosmic noon (z=0.3-1). 
    These are actively growing (orange arrows), and may have originated 
    a billion years ago from the scaling relation (big orange arrow). 
    These sources could be the missing link (dotted gray line) to outliers 
    of local scaling relations, NGC4486B and NGC1277, which feature $M_\bullet/M_\star$ ratios of 3-10\%.
    Galaxy mergers and self-regulation would regress the $M_\bullet/M_\star$ (gray arrows)
    to typical ratios (0.5\%). 
    Our systems are similar to those in the early Universe
    \citep[right-hand side][]{Stone2024,Ubler2023,Harikane2023,Maiolino2024,Yue2024,Juodzbalis2026},
    but we find the highest cosmic space density (see top panel),
    suggesting persistent creation of OMBHs by accretion over cosmic time.
    Grey symbols \citep{Merloni2010,Suh2020,Li2023,Mountrichas2023,Sun2025} cover either higher redshifts (z=1-3) than our sample or not as high $M_\bullet/M_\star$ ratio sources. Blue boxes indicate low redshift works \citep{Reines2015,Li2021,Zhang2023} finding moderate $M_\bullet/M_\star$ ratios.
    Redshifts are illustrative only. We focus the literature comparison on $M_\star>10^9\,M_\odot$.
    }
    \label{fig:illustration}
\end{figure*}

\subsection{A high space density}\label{sec:resultsdensity}
A main result of this work, the abundance of the OMBH population, is presented in the upper panel of Figure~\ref{fig:illustration}.
The systematic survey defining our parent sample allows quantifying the cosmic 
occurrence rate of these systems, which is at least $8$ per 
$0.5\,\mathrm{Gpc}^{-3}$ (purple point in the top panel of Figure~\ref{fig:illustration}), adopting the cosmological volume to redshift 1 of the 
\ero{} eFEDS survey. This space density is higher than the 
results of \cite{Li2021}, who found 10 over-massive systems 
in the approximately 1200\,deg$^2$ HSC-WIDE survey in the $z=0.2-0.8$ 
redshift interval ($\approx4\,\mathrm{Gpc}^{-3}$; 4$\sigma$ significance by a score ratio test \citep{gu2008testing}).
For comparison at similar 
stellar masses, we consider the EIGER JWST follow-up 
\citep{Yue2024,Stone2024} of 6 $z\sim6$ quasars with optical magnitudes 
of $m_\mathrm{1450\AA}<24.4 \mathrm{mag\ AB}$. The space density of such 
luminous quasars is $3\,\mathrm{Gpc}^{-3}$ \citep{Willott2010}, and they 
find half of their follow-up sample (3/6) to be over-massive above 
$M_{\bullet}/M_{\star}>5\%$. While the space densities are uncertain, 
currently we infer a 96 times higher space density (14$\sigma$ significance with the same test) of OMBHs: 3/(0.5\,$\mathrm{Gpc}^3$) vs. (3/6)/(3\,$\mathrm{Gpc}^3$).
Our eight sources allow us to put a conservative, 3$\sigma$ lower limit on the space density of $>4.2\,\mathrm{Gpc}^{-3}$ \citep[Jeffrey-Poisson rate estimate][]{barker2002comparison}.
Considering the scatter in single-epoch masses halves the number density and loosens threshold to $M_{\bullet}/M_{\star}>2\%$ (\cref{sec:resultslauer}), where the final estimate is $8\,\mathrm{Gpc}^{-3}$ with a 3$\sigma$ lower limit of $2.1\,\mathrm{Gpc}^{-3}$. The estimates are listed in Table~\ref{tab:spacedensity}.

Our space density estimate is much higher than previous AGN samples, many at higher redshifts (see top panel of Figure~\ref{fig:illustration}). It is important to stress that all AGN samples give only a lower limit on the OMBH population, of which only a subset indicate their presence as AGN in some wavelengths. Therefore, we can compare efficiencies of selection techniques, but not conclude an evolution over cosmic time from the comparison.

\begin{table}[]
    \caption{Space density constraints on OMBHs in this work, before and after accounting for single-epoch black hole mass scatter. Space densities are $3\sigma$ lower limits.}
    \label{tab:spacedensity}
    \centering
    \begin{tabular}{l l c}
        Constraint & Threshold & Space density \\
        \hline
        \hline
        Observed & $M_{\bullet,\mathrm{SE}}/M_\star>5\%$ & $>4.2\,\mathrm{Gpc}^{-3}$ \\
        Intrinsic & $M_{\bullet,\mathrm{intr}}/M_\star>2.5\%$ & $>2.1\,\mathrm{Gpc}^{-3}$  \\
    \end{tabular}
\end{table}

The black point in the top left corner of Figure~\ref{fig:illustration} 
shows the space density of the two galaxies NGC1277 or NGC4486B within a local 100 Mpc radius. This implies an abundant dormant OMBH population
not apparent as AGN, and is not at odds with the minimum OMBH space densities we derive.

Some of our OMBH population may be descendants of high-redshift Universe OMBHs that have recently been reactivated. 
However, the much higher space density makes this explanation currently unlikely for all OMBHs.
Instead, the higher abundance suggests that the channel for creating or maintaining OMBHs has 
remained intact over cosmic time, and may have been active in a non-negligible fraction of galaxies at cosmic noon.

\section{Discussion}\label{sec:discussion}

In this work, we identified eight OMBHs in a 140 deg$^2$ eROSITA X-ray survey field. This implies a higher minimum space density of OMBHs at $z=0.3-1$ than could be established previously with AGN surveys. In this section, we first discuss the reliability and completeness of the selection in \cref{sec:discussion:selreliability}. Next, we discuss evolutionary scenarios that could have created these eight OMBHs. Comparisons to other samples are discussed in terms of selection efficiency \cref{sec:discussion:selefficiency} and mass distributions \cref{sec:discussion:litcomparison}, in particular to the early Universe.

\subsection{Selection reliability and completeness}\label{sec:discussion:selreliability}

Our results suggest that, considering the Lauer bias and selection effect (see \cref{sec:resultslauer}),
the observed distributions are not in conflict with existing scaling relations. 
In this work, we highlight that among rapidly accreting black holes, overmassive ones are not rare (see section~\ref{sec:resultsdensity}).
This suggests that there is an evolutionary process (see next section) capable of maintaining overmassiveness over cosmic time or creating new OMBHs.
Our results empirically bolster the modeling analysis of \cite{Li2024a} suggesting mechanisms unique to high-redshift Universe or 
black hole seeding may not be needed to create OMBHs.

In our simulated Universe of $M_{\bullet}>5\%M_{\star}$ systems, only 1-10\% of them are X-ray detectable. 
This suggests that the total space densities of OMBHs may be one or two orders of magnitude 
higher than the space density of OMBH quasars inferred in this work.
Indeed, the black point in the top left corner of Figure~\ref{fig:illustration}, although based only on two objects, suggests
a local OMBH space density that is two orders of magnitude higher than our AGN-based measurement.

\begin{figure}
\centering
\includegraphics[width=0.99\columnwidth]{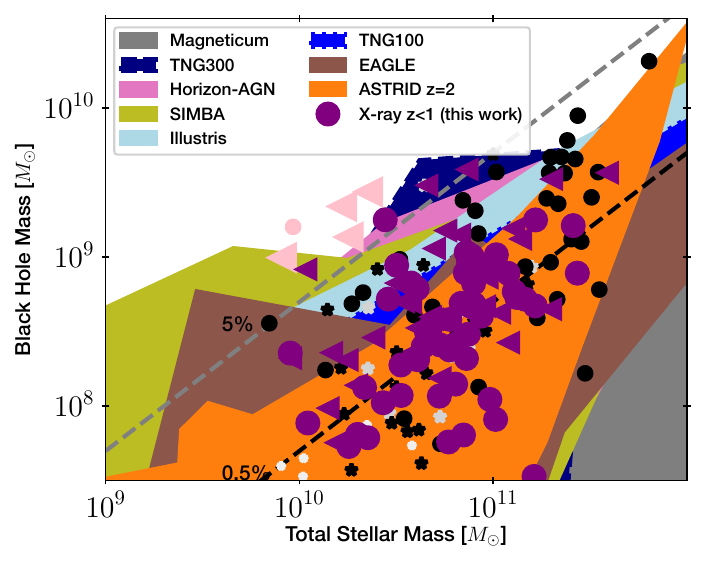}
\caption{\protect\label{fig:litmassessim}
The black hole - stellar mass distribution produced in simulations at z=0 (taken from \cite{Habouzit2021a} for Illustris, TNG, Horizon-AGN, EAGLE and Simba, from \cite{Dolag2025} for Magneticum and \cite{Weller2023} for ASTRID). 
For each simulation, the polygon shows the full range of black hole masses at a given stellar masses
for the redshift z=0 snapshot. As in Figure~\ref{fig:litmassesgal}), our data points are pink and purple symbols, while local galaxies are in black.
Three of our objects (pink left-pointing triangles) lie outside the range of produced of all simulations,
at lower stellar masses at black hole masses of $M_\bullet \gtrsim 10^9 M_\odot$.
}
\end{figure}

\subsection{Evolution scenarios}\label{sec:discussion:evolution}

We can ponder what is needed to create our 
$M_{\bullet}/M_{\star}\sim10\%$ systems at late cosmic times. 
We choose as a starting point the observed $M_{\bullet}-M_{\star}$
relation of quiescent galaxies. 
While a much more massive galaxy could have been stripped of its stars, none of the three OMBH quasars show clear signs of tidal distortions or nearby massive sources, so we leave this hypothesis for future work.
We thus assume the lift above the relations is growth by accretion, which is reasonable as these systems are accreting now, as witnessed by their high X-ray luminosities. 
As a fiducial starting point, we consider $M_\star=10^{10.5}M_\odot$ and assume negligible evolution in $M_\star$.
Thus, we try to identify a BH growth process to grow a $M_{\bullet}\sim10^{8}M_{\odot}$ SMBH 
by $\Delta M_{\bullet}\approx10^{9}M_{\odot}$. 
Because the instantaneous accretion rate may not be representative of an individual system's accretion history, we consider
the typical sample luminosities and the galaxy evolution phase that our \ero{}
selection probes: $L_{\mathrm{0.5-2\,\mathrm{keV}}}\sim10^{43}\,\mathrm{erg/s}$
or $L_{2-10\mathrm{keV}}\sim10^{43.4}\,\mathrm{erg/s}$ \citep{Nandra2025}.
These can be converted into a bolometric luminosity \citep{Marconi2004}
of $L_{\mathrm{bol}}\sim3\times10^{44}\,\mathrm{erg/s}$. Thus,
assuming a radiative efficiency of $\epsilon=10\%$ yields a black hole mass
accretion rate of $\dot{M}\sim4\times10^{7}\,M_{\odot}/\mathrm{Gyr}$.
At constant mass accretion rate, twice the age of the Universe would be needed to 
build such black holes, obviously a nonsensical result.
Instead of constant mass accretion rates, constant Eddington ratios $\lambda$ can be assumed. Then, with the Salpeter e-folding time $\tau_S\approx45 \mathrm{Myr}$, the time to grow by a factor of 10 is $\tau=(\tau_S/\lambda) \ln 10 \approx 0.1\,Gyr$ for $\lambda=100\%$, the maximum of the parent sample reported in \cite{Nandra2025}, and $\tau\approx1\,\mathrm{Gyr}$ for $\lambda=10\%$ (median).
These growth spurts can be interrupted by periods without growth, so that $\tau$
refers to the ``on'' time.
For a scenario where the black hole outgrows the host galaxy mass, 
star formation and stellar mass accretion should be negligible,
thus we assume an isolated galaxy that is not undergoing a star-burst.
During the same time scales, stellar population growth
on the main sequence is negligible at this cosmic epoch \citep{Speagle2014}. 
From this, we conclude that \ero{} identifies 
exponential growth spurts of black holes lasting about a billion years.
Figure~\ref{fig:illustration} illustrates this AGN-driven pathway to 
OMBHs with the large orange arrow. 
Among the non-AGN galaxies in \cite{Kormendy2013}, two OMBHs with ratios
near 10\% stand out, NGC4486B and NGC1277. In this work, we use the latest black hole and stellar mass determinations \citep{Comeron2023,Tahmasebzadeh2025}.
Our systems may be the missing link progenitors to nearby galaxies with such extreme mass ratios of 3-10\% (gray dotted line in Figure~\ref{fig:illustration}).

The channel leading to our OMBHs may apply at other cosmic times, too. 
It is plausible that the JWST-observed OMBH quasars within a Gyr after the Big Bang are created in the same way (right orange arrow in Figure~\ref{fig:illustration}),
starting at mass ratios expected from scaling relations. 
In this scenario, overmassiveness is not due to a high mass at birth, 
but built up after, without lock-step galaxy growth. 
When starting from a seed mass of 1 to 10,000 $M_\odot$, a major observational and theoretical unknown is whether fueling at substantial Eddington accretion ratios can be maintained across the first billion years to transition seeds into the $10^{9-10}M_\odot$ black holes seen as AGN \citep{Inayoshi2020,Volonteri2021,Fan2023}. 
Special circumstances can be at play, as only a minuscule fraction of seeds needs to succeed \citep{Menou2001,Buchner2019b}. 
If the systems found here have started from or below galaxy scaling relations, they may have an analogous, sustained accretion history at substantial Eddington fractions. 
These are worth studying in more detail to shed light on the growth of seeds. 
The constraints in the previous paragraph show that this occurs in at least $4/\mathrm{Gpc}^3$, which we can compare to the cosmic population of massive galaxies ($M_\star>10^{10}M_\odot$) \citep{Ilbert2013}. 
This shows that at least one in a million galaxies succeeds in maintaining such a growth spurt.

At odds with simulations, our results indicate that long excursions from self-regulation are permitted to the AGN and realised in Nature.
Current galaxy evolution simulations can reproduce local scaling relations through calibration \citep{Habouzit2021a,Dolag2025},
but their shape and evolution depend on the assumed black hole feeding and feedback physics.
In virtually all simulations, however, the scatter is narrower than in observations \citep[e.g.][]{Habouzit2021a}.
Figure~\ref{fig:litmassessim} shows the simulations of EAGLE, Horizon-AGN, Illustris and TNG from \cite{Habouzit2021a}, Magneticum \citep{Dolag2025} and ASTRID \citep[at z=2][]{Weller2023}.
Our OMBHs lie outside the range produced in this simulations.
OMBHs akin to NGC4486B and NGC1277 are also not produced in these simulations \citep{Habouzit2021a,Dolag2025,Weller2023}.
This is not primarily because of limited simulation volume, but because the accretion and feedback models tightly link the star-formation rate and black hole accretion with the galaxy's gravitational potential.

A key element in modern galaxy evolution models is feedback from supermassive black holes to suppress star-formation in massive galaxies 
\citep{haehnelt_high-redshift_1998,haehnelt_correlation_2000,kauffmann_unified_2000,volonteri_assembly_2003,wyithe_self-regulated_2003,cattaneo_active_2005,di_matteo_direct_2008,hopkins_cosmological_2008,somerville_semi-analytic_2008,booth_cosmological_2009,croton_simple_2009,sijacki_growing_2009,steinborn_refined_2015,rosas-guevara_supermassive_2016,weinberger_simulating_2017}.
The implemented models vary, but by construction tend to be most effective in the most massive black holes. 
For example, in models where feedback is proportional to the accretion rate, the Eddington limit is less stringent for over-massive black holes than for under-massive black holes. 
A small amount of gas ingested by an OMBH may thus be efficiently converted into a burst of feedback that suppresses star formation galaxy-wide \citep{fabian1999,Granato2004,King2003,Silk1998}, and thus keeping black hole at high $M_\bullet/M_\star$ ratios. 
It is not clear whether, once sufficiently over-massive, black holes can return to the otherwise tight scaling relations.

Even if such systems are the tail of a wider distribution as previously discussed in the literature for JWST \citep{Pacucci2023,Li2024a},
they may hold the key to understanding the conditions for long-term persistent
accretion. Our systems with r band magnitudes of 17-19 AB make detailed investigations feasible. The \ero{} all-sky survey likely holds more than an order of magnitude more of such OMBHs. There may be additional OMBHs in the obscured population, which, while perhaps in the X-ray sample, would need black hole mass measurements other than through optical broad lines.

\subsection{Selection efficiency in comparison with other samples}\label{sec:discussion:selefficiency}

We studied a parent sample of 200 luminous AGN at redshifts $z<1$ and found three galaxies with black holes that are overmassive compared to local galaxy scaling relations. 
In contrast, all 262 type 1 SDSS AGN of \cite{Reines2015} at $z < 0.1$ had a black hole mass to stellar mass ratio ($M_{\bullet}/M_{\star}$) of less than 1\%. The difference between the two groups (three OMBHs versus none) is not statistically significant based on a Wald log-ratio test. Another study with comparable efficiency to ours is \cite{Poitevineau2023}, who report that among the 42 luminous radio AGN, five have $M_{\bullet}/M_{\star}>2\%$. However, they use less reliable SDSS pipeline emission line width measurements, and when inspecting three of the five within the SDSS DR16 catalog of more careful line measurements by \cite{Wu2022}, the black hole masses of the same lines are 0.43, 0.21 and 0.93 dex lower than listed in the original paper. 
The study by \cite{Mezcua2023} of 1161 broad-line selected dwarf galaxies identified seven with unusually massive black holes, but only three have a mass ratio $\sim10\%$. This matches our findings, but their parent sample was five times larger, and therefore the selection efficiency is significantly lower ($p<0.02$, $2.06\sigma$, with the same test as above).
Our high success rate of identifying OMBHs may be due to high effective luminosity threshold on the parent AGN sample combined with high spectroscopic completeness. 

Our parent eFEDS-hard sample, unlike type-1 samples that focus on unobscured AGN, includes both unobscured and obscured AGNs with column densities up to $N_{\mathrm{H}}\sim10^{23.5}\,\mathrm{cm^{-2}}$ \citep{Liu2021a,Waddell2024hard}. Our report of three detections of $M_{\bullet}/M_{\star}>5\%$ is thus noteworthy when compared to type-1 samples that require broad lines. 
The recent X-ray based work by \cite{Tanaka2025} identified only one OMBH in their $<1\,\mathrm{deg}^2$ survey areas, while our $140\,\mathrm{deg}^2$ survey allowed us to put constraints on number densities. Due to our conservative cuts and observing limitations, our parent AGN sample (as any other sample) may contain more OMBHs yet to be identified.

\subsection{Comparison with Other AGN Samples}\label{sec:discussion:litcomparison}

In this section, we compare our findings with other AGN samples, starting at higher redshifts. 
The orange markers in Figure~\ref{fig:litmasseshighz} are a compilation from the SHELLQ sample
\citep{Matsuoka2016,Silverman2025} and JWST \citep{Harikane2023,Ding2023,Ubler2023,Maiolino2024,Stone2024,Yue2024,Juodzbalis2026}. 
We also have comparable stellar-mass upper limits (empty symbols) to those found by \cite{Stone2024}. 
These samples have indicated systems with high black hole to stellar mass ratios (above 10\%), raising questions about 
their growth and formation origins.
Our sample at $z=0-1$ (purple markers) closely resembles the distribution of these high-redshift samples. 

The far-infrared AGN sample by \cite{Dong2016} (red crosses in Figure~\ref{fig:litmasses}) also shares this distribution, 
revealing an OMBH population at redshifts 0.5-3.5. 
Our median black hole mass is above the average in SDSS type-1 selections at comparable redshifts \citep{Li2021a}, 
although both samples share a similar range.
When comparing our AGN sample to the low-redshift Universe, there are similarities in distribution to local elliptical galaxies 
(black dots in Figure\,\ref{fig:litmassesgal}). Late-type galaxies show higher masses due to extra components like spiral arms \citep{Kormendy2013,Greene2019}. Samples favoring smaller black holes, such as SDSS type 1 AGN and water megamasers \citep[green diamonds;][]{Greene2016} appear lower down in Figure\,\ref{fig:litmasses}. 
Our sparse data here suggest that only massive black holes pass the X-ray flux threshold due to power-law Eddington ratio distributions.

The overall distribution of stellar mass and black hole mass depends on sample selection. 
A high flux limit restricts samples to luminous AGNs, partly explaining differences in black hole size between our sample and those observed by JWST. 
\cite{Pacucci2023} reported elevated JWST $z>6$ black holes, but \cite{Li2024a} found the differences could be due to selection biases rather than intrinsic discrepancies. 
Such selection effects are also captured by our simulations (\cref{sec:method:lauer,sec:resultslauer}).
Interestingly, different selection methods (broad optical lines, infrared, X-ray) find similar distributions, 
suggesting that OMBHs are not linked uniquely to one phase of AGN obscuration \citep{Hopkins2008}. 
This finding is limited to moderate extinction levels because broad lines are required for the determination of black hole masses.
Our results suggest that high-redshift conditions or unique black hole seeding processes are not required to account for OMBHs \citep{Stone2024}.

\section{Conclusions}
We present a survey that revealed eight OMBHs with $M_\bullet/M_\star>5\%$, ten times higher black hole masses than the typical ratio in elliptical galaxies. 
Compared to other studies, these black holes stand out:
\begin{itemize}
\item We estimate a high space density of OMBH AGN of ${16}_{-6.4}^{+9.7}/\mathrm{Gpc}^3$. 
\item Our yield of identifying OMBHs, 8 in 140 deg$^2$, is highest compared to any previous AGN survey, thanks to high spectroscopic completeness. 
\item Our simulations suggest that the total space density of OMBHs is at least an order of magnitude higher.
\item For reliable stellar masses $M_\star$, we use the GRAHSP SED-fitting code, which is uniquely unbiased to the presence of a bright, unobscured AGN.
We also introduce a new method for incorporating surely spatially extended flux from photometry data as a lower limit on the galaxy component in the SED fitting, without incurring the systematic uncertainties of parametric image analysis.
\item For reliable black hole masses $M_\bullet$, we use the most reliable of the single-epoch estimators, $H\beta$. 
The effect of scatter from single-epoch $M_\bullet$ estimation and flux limit is quantified with simulations, 
indicating that at least half of our objects are intrinsically at $M_\bullet/M_\star>2\%$.
\item Comparable OMBHs cannot be found in current cosmological simulations.
\item A plausible evolutionary process is exponential growth by sustained gas supply to the black hole over a billion years. Super-Eddington growth is not required, and the current accretion luminosities are below their Eddington luminosity.
\item We speculate that such asynchronous growth should be considered in the first billion years as a process that contributes to lifting black holes well above the relation.
\end{itemize}
Several of our black hole mass measurements are from lines with slight asymmetries and boxy shapes. Future reverberation mapping campaigns (such as SDSS-V) may reveal whether line shape information can refine mass estimates or its uncertainties. Future work will focus on characterizing these low-redshift OMBH quasars and their evolution in more detail.

\bibliographystyle{aa}
\bibliography{ms}

\begin{appendix}

\begin{acknowledgements}
We thank the anonymous referee for helpful comments that improved the manuscript.
JB thanks Benny Trakhtenbrot and Jessie C. Runnoe for insightful conversations.
VNB gratefully acknowledges support through the European Southern
Observatory (ESO) Scientific Visitor Program.
This research was supported by the Munich Institute for Astro-, Particle and BioPhysics (MIAPbP) which is funded by the Deutsche Forschungsgemeinschaft (DFG, German Research Foundation) under Germany´s Excellence Strategy – EXC-2094 – 390783311.
RJA was supported by FONDECYT grant number 1231718 and by the ANID BASAL project FB210003. 
FEB acknowledges support from ANID-Chile BASAL CATA FB210003, FONDECYT Regular 1241005,
and Millennium Science Initiative, AIM23-0001.

This research uses services or data provided by the Astro Data Lab at NSF's National Optical-Infrared Astronomy Research Laboratory. NOIRLab is operated by the Association of Universities for Research in Astronomy (AURA), Inc., under a cooperative agreement with the National Science Foundation. This research has made use of the SIMBAD database \cite{Wenger2000} and the VizieR catalogue access tool \cite{Ochsenbein2000vizier}, operated at CDS, Strasbourg, France.
Software used includes astropy \citep{AstropyCollaboration2013,AstropyCollaboration2018}, topcat \citep{topcat}, stilts (\url{https://www.star.bris.ac.uk}), ultranest \citep{Buchner2021}, matplotlib \citep{matplotlib}, scipy \citep{scipy}, statsmodels \citep{seabold2010statsmodels} and WebPlotDigitizer\footnote{\url{https://apps.automeris.io/wpd4/}}.

This work is based on data from eROSITA, the soft X-ray instrument aboard SRG, a joint Russian-German science mission supported by the Russian Space Agency (Roskosmos), in the interests of the Russian Academy of Sciences represented by its Space Research Institute (IKI), and the Deutsches Zentrum für Luft- und Raumfahrt (DLR). The SRG spacecraft was built by Lavochkin Association (NPOL) and its subcontractors and is operated by NPOL with support from the Max Planck Institute for Extraterrestrial Physics (MPE). The development and construction of the eROSITA X-ray instrument was led by MPE, with contributions from the Dr. Karl Remeis Observatory Bamberg \& ECAP (FAU Erlangen-Nuernberg), the University of Hamburg Observatory, the Leibniz Institute for Astrophysics Potsdam (AIP), and the Institute for Astronomy and Astrophysics of the University of Tübingen, with the support of DLR and the Max Planck Society. The Argelander Institute for Astronomy of the University of Bonn and the Ludwig Maximilians Universität Munich also participated in the science preparation for eROSITA.

The UKIDSS source tables were provided by the WIde Field Astronomy Unit (WFAU) of the University of Edinburgh and originate from UKIDSS DR11 of the WFCAM science archive \citep{Hambly2008WFCAM}.
The UKIDSS project is defined in \citep{Lawrence2007UKIDSS}. UKIDSS uses the UKIRT Wide Field Camera (WFCAM; \citep[WFCAM;][]{Casali2007UKIRT} and a photometric system described in \cite{Hewett2006UKIRT}.
The VISTA Hemisphere Survey is based on observations collected at the European Organisation for Astronomical Research in the Southern Hemisphere at the La Silla or Paranal Observatories under programme ID(s) 179.A-2010(A), 179.A-2010(B), 179.A-2010(C), 179.A-2010(D), 179.A-2010(E), 179.A-2010(F), 179.A-2010(G), 179.A-2010(H), 179.A-2010(I), 179.A-2010(J), 179.A-2010(K), 179.A-2010(L), 179.A-2010(M), 179.A-2010(N), 179.A-2010(O).
This publication makes use of data products from the Two Micron All Sky Survey, which is a joint project of the University of Massachusetts and the Infrared Processing and Analysis Center/California Institute of Technology, funded by the National Aeronautics and Space Administration and the National Science Foundation.

The Legacy Surveys consist of three individual and complementary projects: the Dark Energy Camera Legacy Survey (DECaLS; Proposal ID \#2014B-0404; PIs: David Schlegel and Arjun Dey), the Beijing-Arizona Sky Survey (BASS; NOAO Prop. ID \#2015A-0801; PIs: Zhou Xu and Xiaohui Fan), and the Mayall z-band Legacy Survey (MzLS; Prop. ID \#2016A-0453; PI: Arjun Dey). DECaLS, BASS and MzLS together include data obtained, respectively, at the Blanco telescope, Cerro Tololo Inter-American Observatory, NSF’s NOIRLab; the Bok telescope, Steward Observatory, University of Arizona; and the Mayall telescope, Kitt Peak National Observatory, NOIRLab. Pipeline processing and analyses of the data were supported by NOIRLab and the Lawrence Berkeley National Laboratory (LBNL). The Legacy Surveys project is honored to be permitted to conduct astronomical research on Iolkam Du’ag (Kitt Peak), a mountain with particular significance to the Tohono O’odham Nation.

NOIRLab is operated by the Association of Universities for Research in Astronomy (AURA) under a cooperative agreement with the National Science Foundation. LBNL is managed by the Regents of the University of California under contract to the U.S. Department of Energy.

This project used data obtained with the Dark Energy Camera (DECam), which was constructed by the collaboration of the Dark Energy Survey (DES). Funding for the DES Projects has been provided by the U.S. Department of Energy, the U.S. National Science Foundation, the Ministry of Science and Education of Spain, the Science and Technology Facilities Council of the United Kingdom, the Higher Education Funding Council for England, the National Center for Supercomputing Applications at the University of Illinois at Urbana-Champaign, the Kavli Institute of Cosmological Physics at the University of Chicago, Center for Cosmology and Astro-Particle Physics at the Ohio State University, the Mitchell Institute for Fundamental Physics and Astronomy at Texas A\&M University, Financiadora de Estudos e Projetos, Fundacao Carlos Chagas Filho de Amparo, Financiadora de Estudos e Projetos, Fundacao Carlos Chagas Filho de Amparo a Pesquisa do Estado do Rio de Janeiro, Conselho Nacional de Desenvolvimento Cientifico e Tecnologico and the Ministerio da Ciencia, Tecnologia e Inovacao, the Deutsche Forschungsgemeinschaft and the Collaborating Institutions in the Dark Energy Survey. The Collaborating Institutions are Argonne National Laboratory, the University of California at Santa Cruz, the University of Cambridge, Centro de Investigaciones Energeticas, Medioambientales y Tecnologicas-Madrid, the University of Chicago, University College London, the DES-Brazil Consortium, the University of Edinburgh, the Eidgenossische Technische Hochschule (ETH) Zurich, Fermi National Accelerator Laboratory, the University of Illinois at Urbana-Champaign, the Institut de Ciencies de l’Espai (IEEC/CSIC), the Institut de Fisica d’Altes Energies, Lawrence Berkeley National Laboratory, the Ludwig Maximilians Universitat Munchen and the associated Excellence Cluster Universe, the University of Michigan, NSF’s NOIRLab, the University of Nottingham, the Ohio State University, the University of Pennsylvania, the University of Portsmouth, SLAC National Accelerator Laboratory, Stanford University, the University of Sussex, and Texas A\&M University.

BASS is a key project of the Telescope Access Program (TAP), which has been funded by the National Astronomical Observatories of China, the Chinese Academy of Sciences (the Strategic Priority Research Program “The Emergence of Cosmological Structures” Grant \#XDB09000000), and the Special Fund for Astronomy from the Ministry of Finance. The BASS is also supported by the External Cooperation Program of the Chinese Academy of Sciences (Grant \#114A11KYSB20160057), and the Chinese National Natural Science Foundation (Grant \#12120101003, \#11433005).

The Legacy Survey team makes use of data products from the Near-Earth Object Wide-field Infrared Survey Explorer (NEOWISE), which is a project of the Jet Propulsion Laboratory/California Institute of Technology. NEOWISE is funded by the National Aeronautics and Space Administration.

The Legacy Surveys imaging of the DESI footprint is supported by the Director, Office of Science, Office of High Energy Physics of the U.S. Department of Energy under Contract No. DE-AC02-05CH1123, by the National Energy Research Scientific Computing Center, a DOE Office of Science User Facility under the same contract, and by the U.S. National Science Foundation, Division of Astronomical Sciences under Contract No. AST-0950945 to NOAO.

The Hyper Suprime-Cam (HSC) collaboration includes Japan, Taiwan's astronomical communities, and Princeton University. The HSC instrumentation and software were developed by the National Astronomical Observatory of Japan(NAOJ), the Kavli Institute for the Physics and Mathematics of the Universe (Kavli IPMU), the University of Tokyo, the High Energy Accelerator Research Organization (KEK), the Academia Sinica Institute for Astronomy and Astrophysics in Taiwan (ASIAA), and Princeton University. Funding was contributed by the FIRST program from the Japanese Cabinet Office, the Ministry of Education, Culture, Sports, Science and Technology (MEXT), the Japan Society for the Promotion of Science (JSPS), Japan Science and Technology Agency (JST), the Toray Science Foundation, NAOJ, Kavli IPMU, KEK,ASIAA, and Princeton University.

This paper makes use of software developed for Vera C. Rubin Observatory. We thank the Rubin Observatory for making their code available as free software at http://pipelines.lsst.io/.

This paper is based on data collected at the Subaru Telescope and retrieved from the HSC data archive system, which is operated by the Subaru Telescope and Astronomy Data Center (ADC) at NAOJ. Data analysis was in part carried out with the cooperation of Center for Computational Astrophysics (CfCA), NAOJ. We are honored and grateful for the opportunity of observing the Universe from Maunakea, which has the cultural, historical and natural significance in Hawaii. 

Funding for the Sloan Digital Sky Survey V has been provided by the Alfred P. Sloan Foundation, the Heising-Simons Foundation, the National Science Foundation, the National Aeronautics and Space Administration, the US Department of Energy, the Japanese Monbukagakusho, and the Max Planck Society, and the Participating Institutions. SDSS acknowledges support and resources from the Center for High-Performance Computing at the University of Utah. SDSS telescopes are located at Apache Point Observatory, funded by the Astrophysical Research Consortium and operated by New Mexico State University, and at Las Campanas Observatory, operated by the Carnegie Institution for Science. The SDSS web site is \url{https://www.sdss.org}.

The SDSS-IV is managed by the Astrophysical Research Consortium (ARC) for the Participating Institutions. The Participating Institutions are The University of Chicago, Fermilab, the Institute for Advanced Study, the Japan Participation Group, The Johns Hopkins University, Los Alamos National Laboratory, the Max-Planck-Institute for Astronomy (MPIA), the Max-Planck-Institute for Astrophysics (MPA), New Mexico State University, University of Pittsburgh, Princeton University, the United States Naval Observatory, and the University of Washington. 
SDSS-V is managed by the Astrophysical Research Consortium for the Participating Institutions of the SDSS Collaboration, including Caltech, The Carnegie Institution for Science, Chilean National Time Allocation Committee (CNTAC) ratified researchers, The Flatiron Institute, the Gotham Participation Group, Harvard University, Heidelberg University, The Johns Hopkins University, L'Ecole polytechnique f\'{e}d\'{e}rale de Lausanne (EPFL), Leibniz-Institut f\"{u}r Astrophysik Potsdam (AIP), Max-Planck-Institut f\"{u}r Astronomie (MPIA Heidelberg), Max-Planck-Institut f\"{u}r Extraterrestrische Physik (MPE), Nanjing University, National Astronomical Observatories of China (NAOC), New Mexico State University, The Ohio State University, Pennsylvania State University, Smithsonian Astrophysical Observatory, Space Telescope Science Institute (STScI), the Stellar Astrophysics Participation Group, Universidad Nacional Aut\'{o}noma de M\'{e}xico, University of Arizona, University of Colorado Boulder, University of Illinois at Urbana-Champaign, University of Toronto, University of Utah, University of Virginia, Yale University, and Yunnan University.

\end{acknowledgements}

\section{Higher-resolution imaging}
\label{sec:HSCimages}
\begin{figure}
    \centering
    \includegraphics[width=\linewidth]{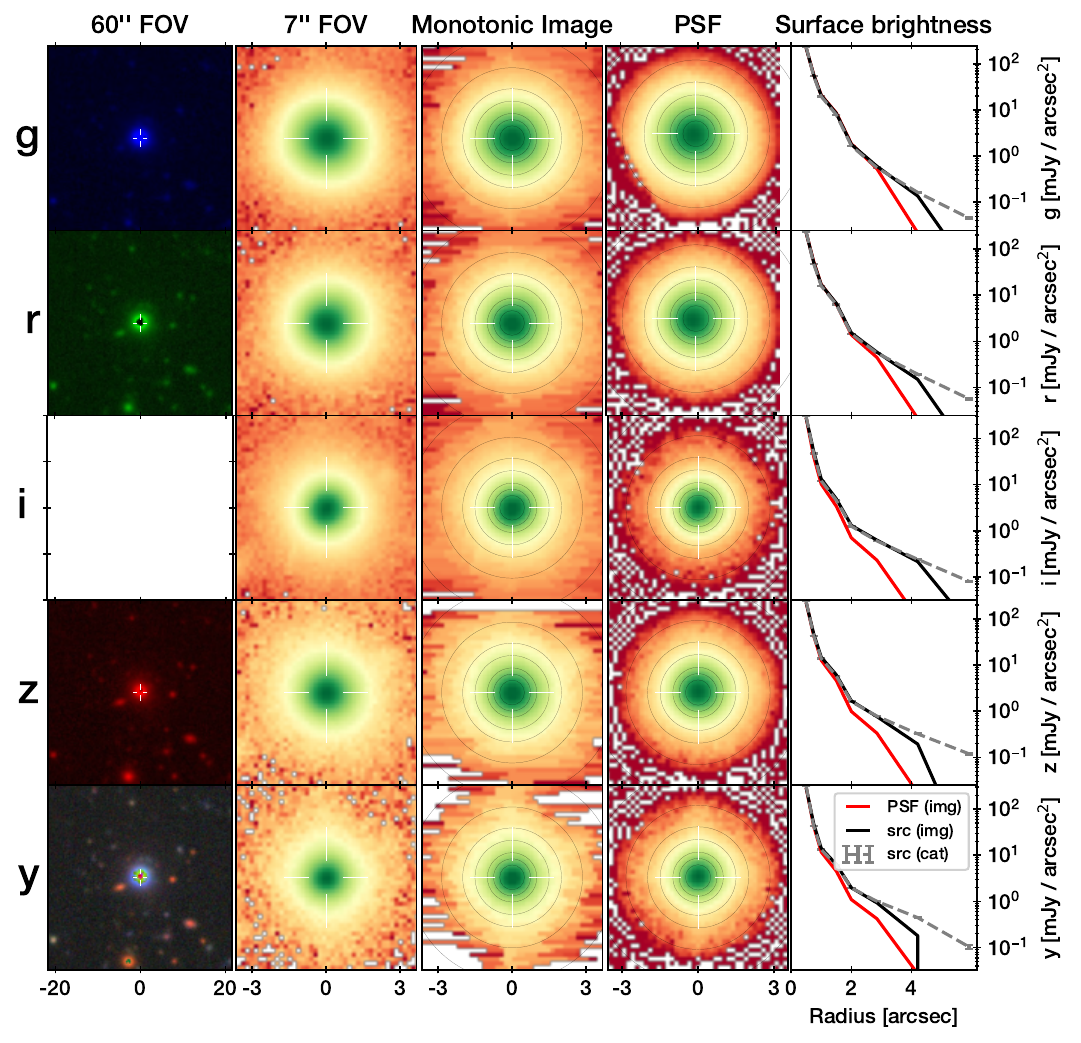}
    \caption{HyperSupremeCam high-resolution images of ID 32.
    Each row represents a filter. The first column shows (only for illustration) shows a wide cutout downloaded from the LS10 server (for y, the grz RGB image is shown).
    The second column shows a 7.2''-wide image cut-out with a logarithmic color scale. White areas have negative flux. The third column shows a cleaned image after enforcing a monotonic decline from the center. The circles indicating apertures of diameters 1, 1.5, 2, 3, 4, 5.7, 8.4, and 11.8''. 
    The fourth column is the PSF image.
    The fifth column shows catalog aperture fluxes (gray error bars and dashed), cleaned image (black) and PSF  (red) matched in normalisation at the center. That the black curve exceeds the red curve indicates the source is extended.
    }
    \label{fig:hsc32}
\end{figure}

\begin{figure}
    \centering
    \includegraphics[width=\linewidth]{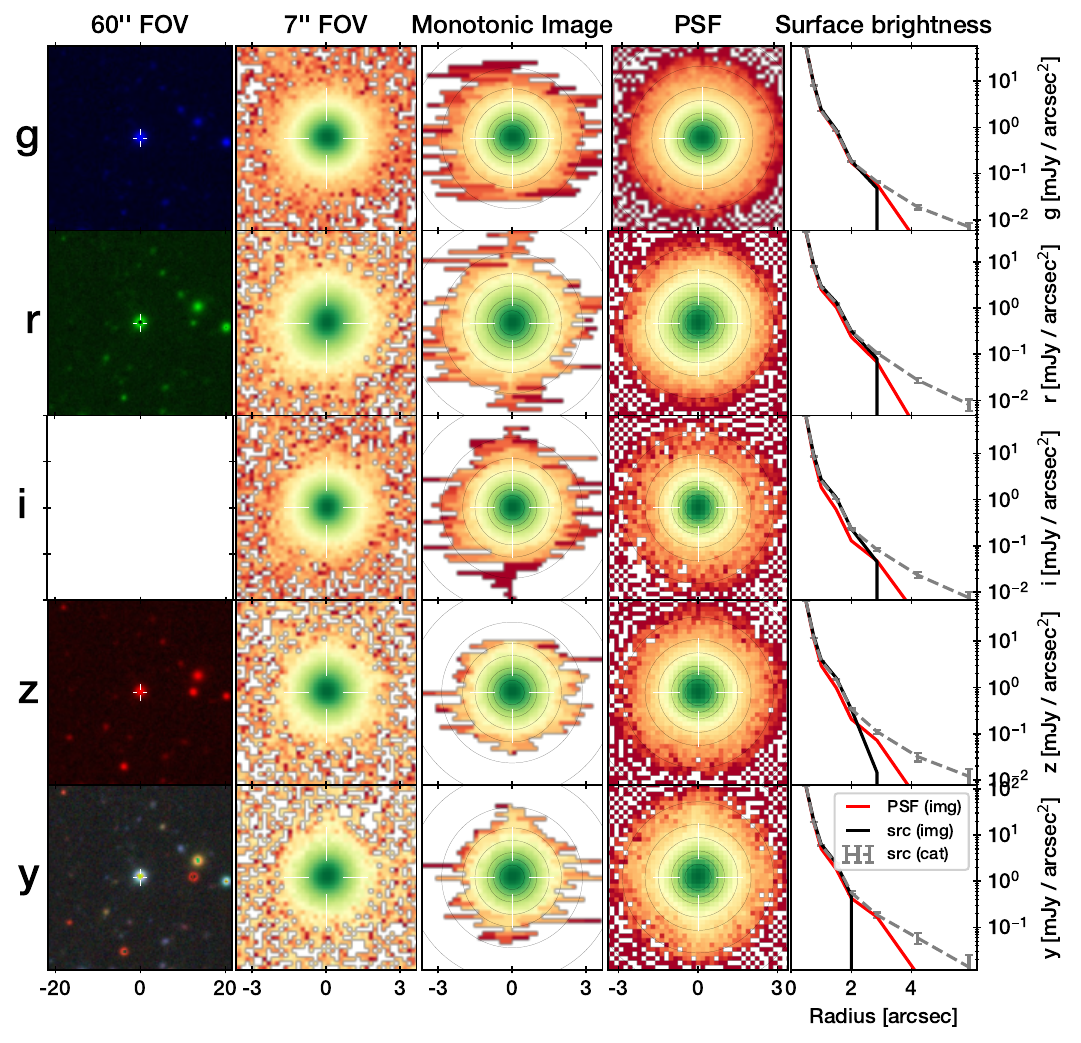}
    \caption{Same as Fig.~\ref{fig:hsc32}, but for ID 133.}
    \label{fig:hsc133}
\end{figure}

\begin{figure}
    \centering
    \includegraphics[width=\linewidth]{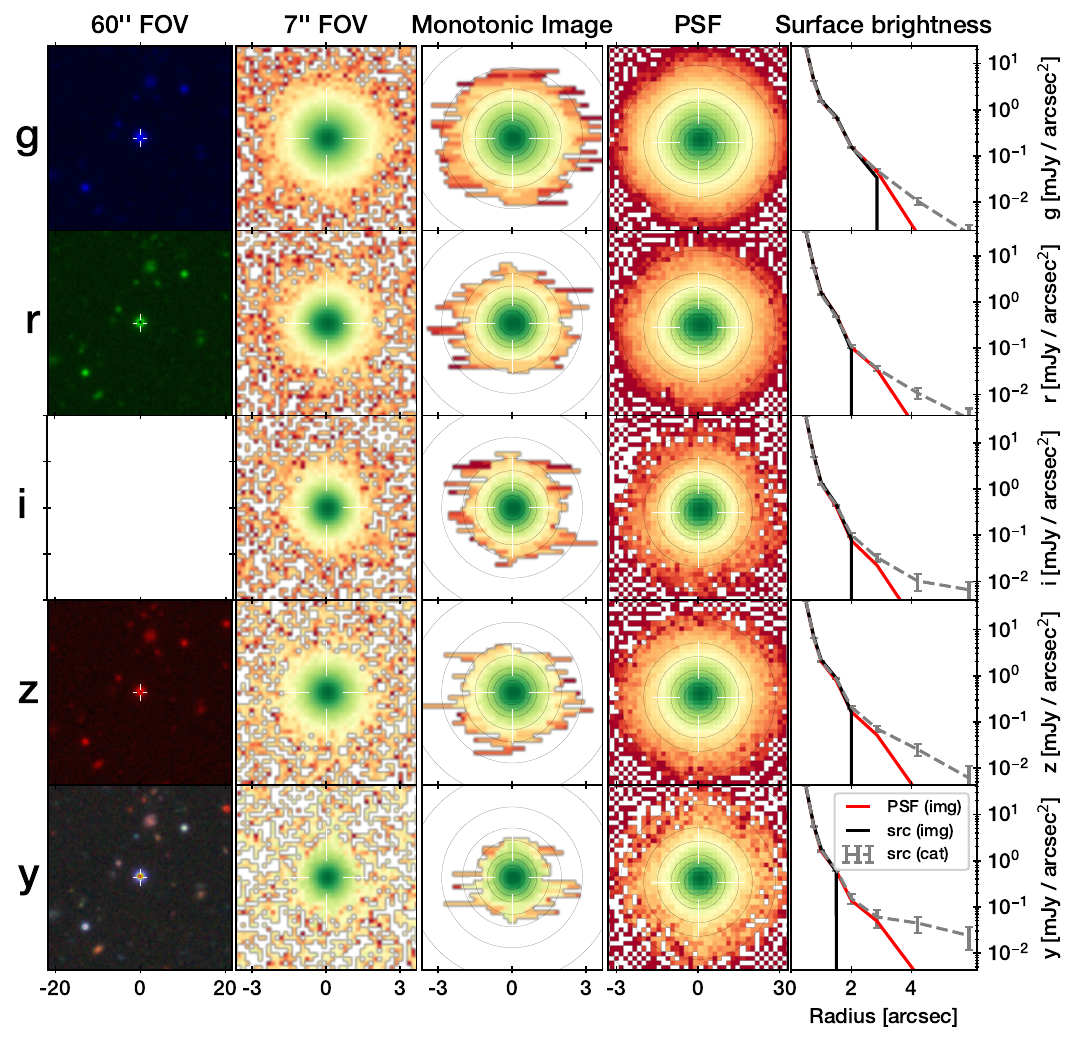}
    \caption{Same as Fig.~\ref{fig:hsc133}, but for ID 1136.}
    \label{fig:hsc1136}
\end{figure}

\begin{figure}
    \centering
    \includegraphics[width=\linewidth]{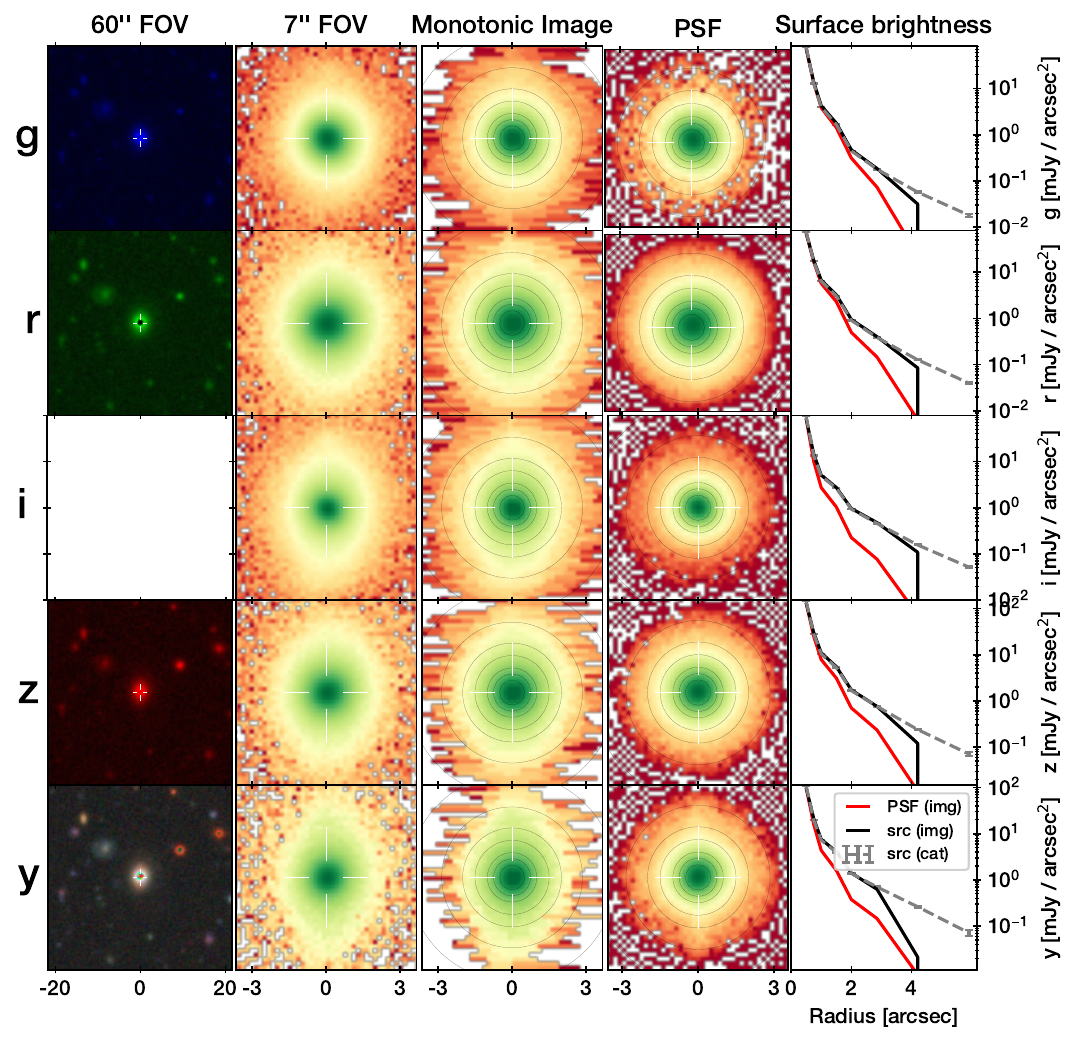}
    \caption{Same as Fig.~\ref{fig:hsc133}, but for ID 45.  The image shows an extended host galaxy. The black radial profile is well above the PSF expectation (red curve).}
    \label{fig:hsc45}
\end{figure}

The eFEDS area has also benefited from observations by the HyperSupremeCam (HSC) survey.
Image cut-outs of our overmassive eFEDS-hard systems are shown in Fig.~\ref{fig:hsc32}, \ref{fig:hsc133}, \ref{fig:hsc1136}, and \ref{fig:hsc45}. We use a floodfill algorithm to enforce monotonic flux decrease from the center, to avoid confusion by unrelated neighbours. We search for an excess of the source surface brightness in comparison to the PSF image surface brightness matched in the inner-most annulus. This non-parametric analysis suggests extended galaxy emission for ID 32, and for the additional candidaste ID 45. ID 133 and 1136 are point-like.

HSC imaging is not used in the main analysis because the long exposures of HSC optimal for studies of the distant Universe cause many eROSITA sources to saturate \cite[see][]{Salvato2021}. Additionally, HSC is not available over the entire eFEDS area \cite[see][]{Salvato2021}. We will analyse the HSC images in more detail in an upcoming paper.

\end{appendix}

\end{document}